# Human Information Processing with the Personal Memex

## ISE 5604
## Fall 2005

## December 6, 2005

| Ingrid Burbey | iburbey@vt.edu |
|---------------|----------------|
| Gyuhyun Kwon | ghkwon@vt.edu |
| Uma Murthy | umurthy@vt.edu |
| Nicholas Polys | npolys@vt.edu |
| Prince Vincent | princev@vt.edu |

# Table of Contents

# 1   Introduction

After World War II, presidential science advisor Vannevar Bush envisioned a future where a "Memex" device would categorize and organize information for the professional (Bush, 1996). For this project, we will explore the Human Information Processing aspects of a Personal Memex, which is a Memex to organize personal (instead of professional) information.  We will consider the use of the Personal Memex by three populations: people with Mild Cognitive Impairment (MCI), those diagnosed with Macular Degeneration and a high-functioning population.  The goal of this study is threefold:

- An annotated bibliography for HIP, usability and design literature relating to the Memex and our populations under study,

- HIP-centered design guidelines for the interface, and

- A low-fidelity prototype.

## 1.1  SenseCam

With the increasing capacity of digital storage and the shrinking sizes of devices, it is now possible to create a device to record our surroundings as we go through our day.  Microsoft Research has developed the SenseCam (Microsoft.com, 2004) for just this use.  The SenseCam, a badge-sized digital camera with light, heat and position sensors, is worn around the neck like a large necklace. Software in the SenseCam monitors the sensors and determines when to snap a picture.  The images and the sensor information are stored in a database.  Gordon Bell, a researcher with Microsoft, and creator of the MyLifeBits project (Microsoft) has worn his SenseCam for two years, recording his daily life.  The MyLifeBits project focuses on organizing, retrieving and displaying the vast amount of information recorded.

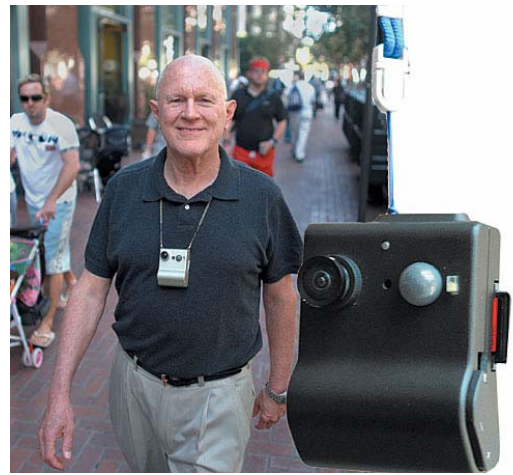

**Figure 1-1 Gordon Bell wears his SenseCam (From (Cherry, 2005))**

## 1.2  Memex

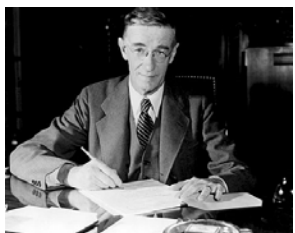

**Figure 1-2 Vannevar Bush (MPI/GETTY IMAGES From (Cherry,**

The idea of a modern device that would record one's life was first proposed by Vannevar Bush in 1945 (Bush, 1996).  He described a desk-sized device which would contain all of one's books, records, correspondence and stored images.  Associative indexing would allow the user to retrieve information and quickly access associated material, in a similar method to the Internet of today.

Many researchers are studying different aspects of how we can use the vast amounts of information that is stored by our computers or in the future, Memex-like devices.   The Remembrance Agent (Rhodes, 1997) watches what you read or write, finds the key words, and displays the relevant parts of your documents, email or web pages.  Lifestreams (Fertig, Freeman, & Gelernter, 1996) displays stored documents as a timeline.  The Haystack Project (Karger & Quan, 2004) is a general purpose information managements system that allows users to link between different types of information, such as emails, documents, photo albums, or appointments, for the purpose of letting "people manage their information in ways that make the most sense to them."

Finding information that has been viewed before is an important aspect of systems that record a person's life.  The Keeping Found Things Found project (Jones, 2005) studies how people organize and re-find their information, such as Web pages (Jones, Bruce, & Dumais, 2001) (also (Capra & Perez-Quinones, 2005) or documents which are part of large projects (Jones, Phuwanartnurak, Gill, & Bruce, 2005).  Stuff I've Seen (Dumais et al., 2003) provides an interface for a desktop user to re-find information such as e-mails, e-mail attachments, files, web pages or appointments.  It allows the user to filter, sort and change the way the information is displayed.

The research challenge of using these technologies to assist people with disabilities has been discussed.  Researchers in England suggest that building a memory aid for elderly people with short-term memory problems is doable within 5 years (Fitzgibbon & Reiter, 2003).  Others see projects like MyLifeBits as the future in caring for people with disabilities (Dawe et al., 2005).  A system is being built to use the automatic capture features of devices like the SenseCam to capture the many details involved in intervention therapy done with autistic patients (Kientz, Boring, Abowd, & Hayes, 2005).

## 1.3  Problem Statement

While much of the current research on organizing and displaying Memex-like information focuses on work-related information, our focus is on personal use of the Memex, outside of the office.  In addition to personal uses by the normal population, we also focused on two classes of users: individuals with Macular Degeneration (MD) and individuals with Mild Cognitive Impairment (MCI). This project is not focused on the internal workings of the Memex device.  We assume that all of the processing, data storage, indexing and retrieval is done efficiently in the background and that any of the data stored can be quickly and easily accessed.

The list of possible applications for the Personal Memex outside of the office is vast and probably incomplete.  We initially considered many possible uses of the Personal Memex, including reminding a user of a forgotten task, recalling a forgotten memory, monitoring what the user is doing, keeping the user on-task, reporting new associations that the user may not have considered, and allowing the user to create their own customized memory schema.  To reduce the scope of the project to a manageable size, the prototype was restricted to two main functions: customizing the Personal Memex for a particular user and showing how the Personal Memex could be used to recall both the details of a previous conversation and additional information that related to that conversation.

## 1.4  Method

This project consisted of several steps as show in Figure 1-3 below.  We began with a literature search to learn about human information processing principles and the needs of our special populations.  We interviewed four experts on campus who work with students who require assistance and the elderly.  We generated several scenarios demonstrating the possible uses of the Personal Memex.  After choosing one scenario, a task decomposition was performed and an initial low-fidelity prototype was sketched out.

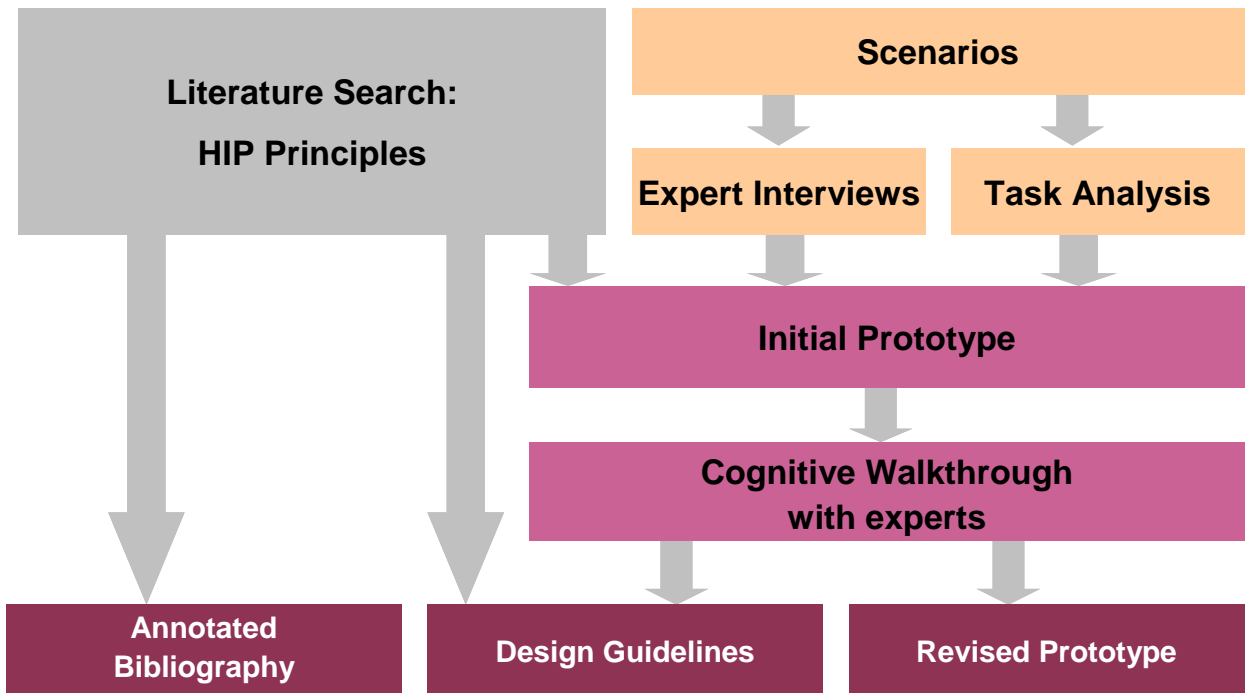

**Figure 1-3 Project Overview**

This project includes deliverables of a set of design guidelines, a low-fidelity prototype and an annotated bibliography.

## 1.5  Outline

This report is organized as follows. Section 2 includes the needs analysis. This includes general human information processing principles that are applicable to the Personal Memex and the unique needs of the populations on which we chose to focus. The interviews with our experts are also summarized in section 2. Section 3 describes the system development portion of our project. It includes our assumptions, a scenario describing one possible use of the Personal Memex, the task decomposition, our initial prototype and the results of the walkthroughs with the experts. Section 4 concludes the report with the results, a final prototype and a list of design guidelines.

This report also includes several appendices.  Appendix A is the original project proposal. Appendix B is an annotated bibliography of the literature researched.  Appendix C details the other scenarios considered in the development of the project and Appendix D contains detailed notes from the interviews held with the experts.

# 2 Needs Analysis

To understand how the Personal Memex system could be used to find and use digital memories, we first examined the literature on relevant issues and principles from Human Information Processing (HIP) for each target population. We then interviewed a number of experts in the field of disabilities and assistive technologies. Each was given a brief system description and questioned about the needs of the populations and necessary features. Finally, we derived a set of design requirements for our prototype.

## 2.1 HIP Principles

### 2.1.1 General Population

We identified the current HIP models that should inform the design of the Personal Memex. This profiling from the literature helps to understand the nature of the problem and collect the requirements for the Personal Memex. First, we examined the nature of *Human Memory,* which should be applicable to general populations of 'high' or 'normal' functioning individuals.

We start with the well-accepted 'Multi-Store Model' of Atkinson & Shiffrin (Atkinson & Shiffrin, 1968), which is shown in Figure 2-1. Here, the distinction is made between Sensory, Working/Short-term, and Long-term Memory. Working Memory includes both processing and storage capabilities to interface between perception, Long-Term Memory, and motor processes. Working memory operates on the content of attention (focused, divided, and switching) and connects with other components such as LTM and motor programs. Long Term Memory, in contrast, is the system responsible for our persistent ('crystallized') memories of procedural, episodic, and semantic information. While the precise nature of how these memory systems are composed and interact is still an open research question, the qualitative distinction is the basis for most memory research.

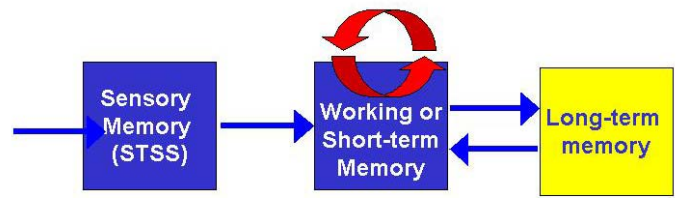

**Figure 2-1 The Multi-Store Model**

### 2.1.2 Long-Term Memory (LTM) - representation and access

The research from Cognitive Psychology into Long Term Memory is concerned with the encoding, storage, and retrieval of memories including learned procedures, episodes and experiences, and knowledge or facts that persist in a person's mind over time. These three different types of LTMs come from studies of Amnesiacs where tasks requiring certain kinds of memory were uniquely disrupted by brain damage or disease (e.g. (Vargha-Khadem et al., 1997)). Indeed a large body of neuroscience evidence is accumulating that implicates different brain regions for the handling of skill, event, and semantic memories (Eysenck & Keane, 2000;

Tulving, 1972). However, the distinction may not be clear in terms of content, but rather the subjective experiences during encoding and retrieval (e.g. episodic requires conscious involvement) (Tulving, 2002).

Descriptive terms for LTMs include explicit (also called declarative) and implicit (non-declarative) memories with the former being verbalisable and accessible to consciousness and the latter not (things like skills, habits, perceptual learning, conditioning, and reflex pathways) (Cohen, 1984). Explicit memories are traditionally measured by free and cued recall tasks and/or recognition tasks.

What are the structures of Long-term memory representations? F.C. Bartlett coined the terms 'Schema/Schemata' in 1933 as "the way in which knowledge in a particular area is held to be organized in memory" (Bartlett, 1995). The attributes of Schemata can be summarized as:

- Developed by experience and pre-programming– accommodation & assimilation
- Prototypical elements are always given priority.
- Can be applied non-consciously
- Shared across individuals within a culture
- Stable over time

Finally, there are some important phenomena related to Memory that must be mentioned. The recall process can be confounded by *pro* and *retro -active interference*. Proactive interference refers to prior learning interfering with subsequent learning; retroactive interference refers to post hoc learning confounding the prior learning (Wickens & Hollands, 2000).

### 2.1.3   Episodic & Autobiographical Memory

Episodic memory refers to the memory of event that occurred sometime in the past. Autobiographical Memory is more specific; it concerns memories for personal events and experiences (M. A. Conway, 1990, 1996; M. A. Conway & Rubin, 1993).  Retroactive interference for event memory is a much-studied subject especially in regards to the validity of eyewitness testimony, suspect identification, leading questions, etc. (Neisser, 1982; Wells & Olson, 2003)

### 2.1.4   Associative Memory - cues

Wiseman and Tulving (Wiseman & Tulving, 1976) originally formulated the Encoding Specificity Principle, which maintains that items are encoded with respect to their context and that retrieval is a function of similarity to that encoding context (Eysenck & Keane, 2000). Originally, this was Tulving's (Tulving, 1972) explanation of Cue-dependent forgetting – the situation where information is stored in LTM but cannot be retrieved sure to inadequate retrieval cues. This principle helps to explain the influence of context in recall and recognition memory performance. The greater the information overlap between the context at encoding and the context at retrieval, the greater the probability of successful retrieval.

Baddeley (Baddeley, 1982) further clarified the role of context by distinguishing between intrinsic and extrinsic context: intrinsic context is the semantic association and significance of a stimuli whereas extrinsic context refers to the environment. From his and colleagues' experiments, intrinsic context affects recognition and recall, but extrinsic context only affects recall.

So context and cues are key factors in LTM performance. In addition, there are other factors that contribute to the durability and robustness of a memory. For example, Craik & Lockhart (Craik & Lockhart, 1972) and Lockhart & Craik (Lockhart & Craik, 1990) propose a Level-of-Processing effect where the deeper the level of analysis and the more attentional resources are dedicated to it during learning, the richer and more persistent the memory trace (i.e. rehearsal).

Cueing and priming to retrieve memories is a crucial ability but may not be simple. Tulving & Schacter (Tulving & Schacter, 1990) proposed that priming is supported by a separate Perceptual Representation System (PRS). Dissociation evidence (age, amnesia, normals) suggests a different processes and brain areas are responsible for priming versus recall. Visual priming is typically an activity with implicit types of memory and both verbal or nonverbal priming types are hyperspecific and relatively inflexible.

### 2.1.5  Working Memory – capacity and control

Working Memory includes both processing and storage capabilities to interface between perception, Long Term Memory, and motor processes. Typically, Short-term Memory refers to the storage buffer while Working Memory refers to the items being processed. As Miller (Miller, 1956) observed, there is a capacity limit to the number of information items that can be operated on at a given time; this 'magic' number ($5 \pm 2$) refers to the number of information 'chunks' that can be processed at a given time. The duration of storage for items in Short-term memory falls between 12 and 25 seconds (Smith-Jackson, 2005; Wickens & Hollands, 2000).

Working memory in this formulation (e.g. (Baddeley, 2003)) consists of multiple fluid components including the visuospatial sketchpad, the episodic buffer, and the phonological loop. These 3 subcomponents are managed by a central executive component that manages attention (focusing, dividing, and switching) and connects the other working memory components to LTM. Support for this model comes from a number of empirical results that demonstrate interference within but not between verbal and visual stimuli. Long Term Memory, in contrast, is the system responsible for our persistent ('crystallized') memories of procedural, episodic, and semantic information. While the precise nature of how these 2 memory systems are composed and interact is still an open research question, the qualitative distinction is the basis for most memory research.

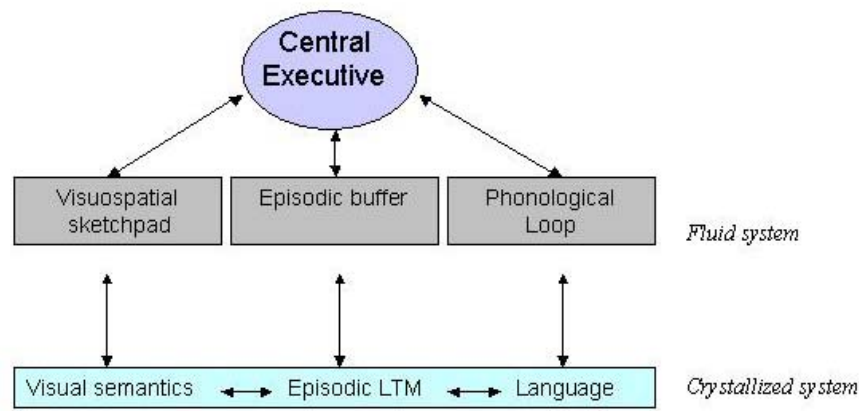

**Figure 2-2 Revised WM Architecture**

Ericsson & Kintsch (Ericsson & Kintsch, 1995) propose an alternate explanation of cognitive architecture that can account for two important facts that they claim Baddeley's modal model of WM cannot. First, experts and skilled performers have expanded memory capacities for complex tasks in their domain. This capacity allows then to efficiently access and use knowledge from their LTM. Second, skilled performers can stop working on a problem and resume it later without a major performance impact. Instead of simply replacing STM with WM (what they call STWM), Ericsson & Kintsch propose expanding the model with another store: Long-Term Working Memory (LTWM), which is a more stable storage medium and is accessible via cues in STWM. Through study and practice in some domain (including text comprehension), these cues can be structured to efficiently encode and retrieve information between STWM and LTM.

Ericsson & Kintsch studied expert and domain-specific strategies in 5 task domains: mental abacus calculation, mental multiplication, dinner orders, medical diagnosis, and chess. In each, they found evidence of distinct retrieval structures being used to avoid interference and improve performance. This notion of mnemonic skill is consistent with Maguire et al's recent work (Macguire, Valentine, Wilding, & Kapur, 2003) of neuroimaging world-class memory performers. According to Ericsson (Ericsson, 2003), the different patterns of fMRI activation between experts and controls is attributable to the differences in memory strategy - experts have established retrieval structures using their imagery and 'method of loci' strategies, strategies the controls did not use. In turn, these expert strategies activated brain areas associated to spatial memory and navigation.

Just & Carpenter (Just, Carpenter, & Keller, 1996) take a different view on WM and make a capacity argument to explain individual differences in task performance. In their theory, there are capacity constraints in WM and the degree of information encapsulation in WM determines the efficiency and power of the system. Thus, performance is determined by the management of the WM chunks within capacity and not by separate modular resources. In their work, they used a reading span test (attributed to WM), dividing subjects into high, medium, and low span groups. Subjects read passages with complex sentences (e.g. subject and object relative clauses) and ambiguous or polysemius words. As predicted for the variety of sentence types, high-span individuals were better are comprehension correctness and seemed to be able to maintain ambiguous terms across long sentences to their semantic resolution. Some researchers have gone so far as to extended this capacity argument to implicate WM as being a principle factor in *g* or general fluid intelligence (A. R. A. Conway, Kane, & Engle, 2003; Miyake, Friedman, Rettinger, Shah, & Hegarty, 2001).

So since the early formulation of STM, alternate models of cognition and memory have emerged including WM, WM with capacity constraints, and the distinction between STWM and LTWM components. How each accounts for individual differences in cognitive and memory performance is at the center of the debate. Baddeley's approach is an architectural one that describes WM and divides subcomponents in order to replace the traditional STM. Ericsson & Kintsch maintain that Baddeley's model is insufficient to account for domain-specific expertise and add another processing module to encode and retrieve information from LTM. Contrasting Ericsson & Kintsch, Just & Carpenter claim that another module is not needed to account for expertise but rather that capacity in WM is the significant issue. At this point it is difficult to speculate are to whether these theories will converge or a winner will take all. While further research is required into the methods of chunking and encapsulation in WM and the nature of central executive functions, we expect the former.

### 2.1.6   Cognitive Load Theory

In terms of the nature of cognitive processing and interface design, we are considering how the Memex can reduce the load between Working Memory and Long Term Memory. This has been addressed via the Cognitive Load Theory of Chandler & Sweller (Chandler & Sweller, 1991). Chandler & Sweller showed that co-references between text and images can improve the comprehension of complex instructional material. Here, the combination of visual and phonological cues provide a more robust encoding of the material to be remembered. Similar results have been shown in multimedia comprehension (e.g. Faraday & Sutcliffe, 1997, 1998) where the integration of text and image information in multimedia presentations has resulted in better user recall than images only.

More recently, Mayer [2002] described how cognitive theory informs multimedia design principles for learning. To evaluate meaningful learning, they use transfer questions including troubleshooting, redesigning, and deriving principles. Using the dual-channel, limited capacity model of Baddeley (above), he suggests that users are actively selecting and organizing mental representations from pictoral and verbal channels and integrating them into a coherent model and relating that integrated model to other knowledge in long-term memory. Because on-screen text can interfere with the visuospatial channel, speech narration is generally better that text. The benefits shown from these investigations are promising to consider for the Memex multimedia interface.

### 2.1.7   Perception

One important Human Information Processing model for perception is called 'Signal Detection Theory'. This model concerns the detection of a signal in 'noisy' conditions (i.e. under uncertainty).  Whether the signal (visual or an auditory) sensation is perceived depends on two factors: beta and d'. beta is a term that describes the subjective level of certainty in the human operator. d' refers to the sensitivity of the sensory system. Both of these factors may be manipulated by design. For example, a visualization or display technology might change the salience of a stimuli to overcome a sensitivity threshold or a decision support tool might guide the operator to explicitly consider certain information or procedures to reduce a risky criterion or bias.

The nature of visual perception is obviously a crucial factor in the design of effective graphics. The challenge is to understand human perceptual dimensions and map data to its display in order that dependent variables can be instantly perceived and processed pre-consciously and in parallel [Friedhoff, 2000].  Such properties of the visual system have been described (ie sensitivity to texture, color, motion, depth) and graphical presentation models have been formulated to exploit these properties, such as pre-attentive processing [Pickett et al, 1996; Triesman & Gormican, 1988] and visual cues and perception [Keller, 1993].

By taking account of how humans build their cognitive models and what perceptual predispositions and biases are in play, designers can take steps to minimize or leverage their effect. This line of inquiry has been termed the 'inverse problem of design' by Joseph Goguen (Goguen, 2000) and 'Information Psychophysics' by Colin Ware (Ware, 2003). Information Psychophysics is one step above Psychophysics, which examines how a sensation becomes a perception;   Information Psychophysics concerns presenting the user with the perceptual substrates for comprehension and insight. Norman (Norman, 1986) proposed a descriptive model for the cognitive engineering of interfaces and identified the distance between user and system as the 'Gulfs' of Evaluation and Execution.  Reducing these gulfs is the principle goal of interface design.

Also related to Memex interface design and the Gulf of Evaluation is the work on specific rankings of information representations. These rankings summarized from comparative graphics are task-specific (Wickens, 2000). The first set of mappings, well known in information visualization, applies to point reading and local comparisons ranks (Table 2-1 below). When the task is integrative - requiring complex comparisons or requiring synthesis of information – the ordering may be reversed, preferring more 'integral, object-like' displays. This is known as the 'Proximity-Compatibility Principle' (Wickens & Carswell, 1995), which describes why there are fewer mental operations required when the spatial arrangement of information sources is compatible with the cognitive integration requirements.

**Table 2-1 Accuracy rankings for point reading and local comparison of visual markers by general data type**

| Data Type | Quantitative | Ordinal | Nominal |
|---|---|---|---|
| Graphical Representation | position<br>length<br>angle / slope<br>area<br>volume<br>color / density<br>　　(Cleveland & McGill, 1984) | position<br>density<br>color<br>texture<br>connection<br>containment<br>length<br>angle<br>slope<br>area<br>volume<br>　(Mackinlay, 1986) | position<br>color<br>texture<br>connection<br>containment<br>density<br>shape<br>length<br>angle<br>slope<br>area<br>volume<br>　(Mackinlay, 1986) |

## 2.1.8  Aging

As people age, a number of functional, cognitive, and behavioral changes occur in regards to their memory and cognitive processing. Neurologic changes associated with aging such as metabolism, blood flow, brain tissue volume, and neurochemistry are thought to set off a cascade of impairments that effect memory performance. According to Anderson & Craik (Anderson & Craik, 2000), these include a reduction in attentional resources, cognitive slowing, and consequently reduced cognitive control. This reduced control manifests as impairments for: prospective memory, inhibition, and conscious recollection.

Using the experimental paradigm of divided attention, it seems that tasks were disrupted more in older adults when the encoding or retrieval aspects were more demanding. Under the environmental support hypothesis (Craik, 1986), the quality of retrieval cues in the environment can offset the demanding nature of memory retrieval and help equalize performance on tasks such as cued recall and associative recognition. This suggests that a supportive task environment can reduce some memory decrements associated with aging. The speed of cognitive operations in aging adults is also reduced (with about 1.5 times the latency as young adults). Both of these factors contribute to the reduction of cognitive control which includes: more distractability from irrelevant information, maintaining non-task-relevant information, inhibiting extraneous information for recall, and interference in recognition (Anderson & Craik, 2000).

Reuter-Lorenz (Reuter-Lorenz, 2002) found compelling evidence that aging may not simply be a progressive decline in cognitive ability. Using fMRI, she studied working memory and episodic memory retrieval in young and old adults. What she found is that older people recruited different parts of the brain for the same tasks; for example older subjects showed that their hemisphere activation was more balanced than the highly lateralized activation in younger subjects. This difference and possible re-localization over time suggests that aging people are reorganizing their cognitive functions and that neural plasticity may continue will throughout the lifespan. The differences in active brain regions also suggests the potential of compensation for age-related memory impairments.

Researchers at UNC have examined the relationship between aging and working memory (Dumas & Hartman, 2003; Hartman, Dumas, & Nielsen, 2001). Hartman et al. used a delayed-matching-to-sample task to investigate the age differences in perception and working memory. They found that older adults are slower to perceive stimuli such as color, location, and order. If stimuli were equated for this difference in perceptual speed, no differences in working memory were found. Finally, they examined the effect of study time (encoding in working memory) on performance for pattern/letter comparison, digit symbol, and letter-sequencing tasks. Older adults were also able to benefit from increased study time, but required longer to achieve this performance parity. This research also suggests that given more time for perception and working memory encoding, older adults can perform as well as younger adults in terms of accuracy in perception and working memory.

### 2.1.9 Macular Degeneration / Visual Impairment

Macular degeneration is a chronic eye disease that occurs when tissues in the macula deteriorates. Macula is the part of the retina that is responsible for central vision. The retina is the layer of tissue on the inside back wall of your eyeball. Degeneration of the macula causes blurred central vision or a blind spot in the center of your visual field. This condition mostly occurs in the older population and hence it is usually called age related macular degeneration in some literature. Macular degeneration is the leading cause of severe vision loss in people age 60 and older. More than 1.6 million American adults have the advanced form of age-related macular degeneration.

The macula is made up of densely packed light-sensitive cells called cones and rods. Cones are essential for central vision as well as color vision. When the light-sensitive cells of the macula (cones) become damaged, it can no longer send normal signals through the optic nerve to your brain, and your vision becomes blurred. This is often the first symptom of macular degeneration.

Macular degeneration usually develops gradually. The signs and symptoms of the disease may vary, depending on which of the two types of macular degeneration you have.

With dry macular degeneration the following symptoms are observed:

- Colors that appear less bright
- Difficulty recognizing faces
- Blurred or blind spot in the center of your visual field combined with a profound drop in your central vision acuity
- Increasing difficulty adapting to low levels of illumination.
- Gradual increase in the haziness of your overall vision.
- Printed words that appear increasingly blurry
- The need for increasingly bright illumination when reading.
- A need to scan your eyes all around an object to provide a more complete image

With wet macular degeneration the following symptoms are observed:

- Decrease in or loss of central vision
- Visual distortions, such as straight lines appearing wavy or crooked, a doorway or street sign that seems out of whack, or objects appearing smaller or farther away than they should
- Central blurry spot

### 2.1.10 Mild Cognitive Impairment

Mild cognitive impairment (MCI) can be considered as a stage between the cognitive changes of normal aging and the more serious problems caused by Alzheimer's disease. There are lots of disagreements related to what actually defines mild cognitive impairment. This is not considered as a stable state. This condition can deteriorate into other types of cognitive disorders.

Mild cognitive impairment can affect many areas of cognition such as language, attention, critical thinking, reading and writing. They also show mild difficulties in other areas of thinking, such as naming objects or people (coming up with the names of things) and complex planning tasks. Mild cognitive impairments can cause a destabilizing effect on the normal lives of people affected by it and can lead to depression and other related conditions.

Mild cognitive impairment can be divided into two broad subtypes. One subtype, amnestic mild cognitive impairment, significantly affects memory while the other type, nonamnestic mild cognitive impairment, does not. Other functions, such as language, attention and visuospatial skills, may be impaired in either type. It has been estimated that amnestic mild cognitive impairment converts to Alzheimer's at a rate of 10 percent to 15 percent a year.

Hippocampus is the area of your brain responsible for processing, storing and recalling new knowledge and information it has a major role in the memory system it's in charge of sorting new information and sending it to other sections of your brain for storage. The hippocampus then recalls information when it's needed. It also makes associations between your newly acquired memory and previously stored memories. In most people, the hippocampus shrinks with age and causes mild memory decline. In people with mild cognitive impairment, this shrinkage is at a faster rate than normal causing memory impairment and other cognitive problems.

Even though, the exact prevalence of mild cognitive impairment in the population is unknown, it is estimated that it affects around 20 percent of the nondemented population over age 65. Only about a third of those cases, however, have the amnestic variety that has been linked to Alzheimer's. Mild cognitive impairment is considered as the leading ailment among the people above 65 years.

Some level of mild memory loss is associated with normal aging. But, when you start forgetting things you typically remember, such as doctor's appointments and if this happens very often, it could be a symptom of amnestic (memory-related) MCI. There are no clear cut symptoms for mild cognitive impairment. Some of the commonly used guidelines a diagnosis of amnestic MCI are:

- Deficient memory, preferably corroborated by another person
- Essentially normal judgment, perception and reasoning skills
- Largely normal activities of daily living
- Reduced performance on memory tests compared to other people of similar age and educational background
- Absence of dementia

## 2.2  Interviews

This project focuses on three user populations: individuals with mild cognitive impairment, individuals diagnosed with Macular Degeneration and individuals without these disabilities. We interviewed several professionals at Virginia Tech who work with students and adults with disabilities. The goal of the interviews was mainly to get a summary of the effects of these disabilities. A complete summary of the interviews, including suggestions for novel uses of the Personal Memex, is included in Appendix D.

Our experts included:
Bill Holbach, Assistive Technologies Coordinator
Hal Brackett, Assistive Technologies Lab and Special Services Manager
Karen Roberto, Professor, Department of Human Development and Director, Center for Gerontology
Virginia Reilly, ADA Coordinator

A brief description of the SenseCam and the idea of the Memex was distributed to each expert along with a short list of questions for discussion.   This description is shown in Table 2-2.

**Table 2-2 Executive Summary for our Interviewees**

| Description of a Personal Memex |
| --- |
| The "Personal Memex" is a device that categorizes and organizes information for a user. It makes use of a SenseCam, which is a 'badge-sized wearable camera' developed by |

Microsoft Corporation. The SenseCam will record, throughout the day, pictures, audio recordings and other sensor data. In addition to information from the SenseCam, the Memex will also make use of other personal information like the user's calendar, phone logs, email records and web browsing history. For our project, we assume that this massive amount of collected information is magically, efficiently organized during the night and any part of it can be easily accessed. The interface to access information organized by the Memex can range from a cell phone, to a laptop PC, to a personal audio system, to a custom-built interface.

Many applications come to mind for such a device, such as remembering people, alerting the user when she forgets to take a left turn on the road, providing details of a previously seen car, or reminding her where she left her keys.

For our project, we will first focus on building the schemas that people use to organize their information. To be truly useful, the Memex will need to organize information using mental maps similar to those of the user. For example, information that the user finds useful, what she likes/dislikes, what she considers important, etc. Thus, mental maps will be based on associations made between pieces of information. Once the Memex organizes information, it can be used to either recall a memory, or remind the user, or even make recommendations based on patterns it sees.

Our initial prototype will be developed for three population groups: those with mild cognitive impairment, those with macular degeneration and those without these disabilities.

## Questions for our Experts

We would like you to consider the following questions as you ponder how a personal Memex would be used by a user with macular degeneration or a user with mild cognitive impairment.

- What are the most important functions for the target population?

- Which features would be nice to have?

- Do you see any obstacles for the target population trying to use such a device?

- Apart from the uses mentioned, do you forsee any other use of this device for a particular type of population?

### 2.2.1 Interview Summaries

#### 2.2.1.1    Bill Holbach and Hal Brackett, Assistive Technologies

Bill Holbach is the Assistive Technologies coordinator and Hal Brackett is the Special Services Manager. The purposes of the Assistive Technologies and Special Services groups at Virginia Tech include providing equal access to technology for all students at the University by providing various computer-related assistive technologies, such as screen magnifiers, voice recognition software and accessible workstations. More details about Assistive Technologies can be found at http://www.it.vt.edu/organization/lt/assistive_technologies.html.

In their role providing services for students with a wide range of disabilities, Bill and Hal have both expertise and anecdotal evidence on what is needed and useful to the student population on campus.

In general, individuals with Macular Degeneration see no color or detail.  The condition is similar to having night vision, and many individuals prefer to keep light levels low. However, Macular Degeneration can range from very little visual impairment to complete blindness.  Even within one individual, the affect of the disability changes over time.  Bill and Hal suggested that we focus on general visual impairment, realizing that no single interface will work for all users, or even a single user over time.

Mild Cognitive Impairment was discussed along with Learning Disabilities and other cognitive impairments.  Individuals with learning disabilities are easily distractible and need to focus. Students with Learning Disabilities need to learn the changes needed in their daily routine to succeed on campus.  Studies have shown that failure in college for students with learning disabilities is not due to a lack of intelligence, but in problems with organization.  As we learned with Macular Degeneration, designing for MCI needs to support a wide range of effects and the design needs to be adaptable to a broad and varied audience.

### 2.2.1.2    Dr. Karen Roberto, Center for Gerontology

Dr. Karen Roberto is a Professor in the Department of Human Development and Director of the Center for Gerontology.  Dr. Roberto's research focuses on health and social support in the later stages of life, so she considered the Personal Memex as used by an older population.

People with cognitive disabilities have difficulty recalling calendar and time events. Relating events to time can be mentally taxing for them. They are more comfortable recalling visual images. MCI patients have a number of problems with time including perception and instruments for management such as calendars. Numbers are problematic. This population can have a tendency to forget certain categories of things including: objects, appointments, recent actions (done something), and names. They can also have difficulties with directions.

Like Bill Holbach and Hal Brackett, Dr. Roberto reported that the effects of MCI can vary greatly and we are realizing the importance of designing customization and adaptability into the device.

### 2.2.1.3    Dr. Virginia Reilly, Office for Equal Opportunity and Affirmative Action

Dr. Reilly is the ADA coordinator for the EO/AA at Virginia Tech.

People with macular degeneration prefer dim lit conditions than brightly lit conditions and high contrast.  There will be lot of differences in designing for this group depending on whether the visual impairment was acquired or if it was present at birth. If it's acquired then the person is going to have a well developed visual schema where as you won't have a developed visual schema in the other case.

As with our other experts, Dr. Reilly stressed that no one design template could be universally used.  A highly, customizable, but easy-to-use interface should be the goal.  Dr. Reilly also reviewed other assistive tools that already exist on the market and remarked that it would be preferable if our design was compatible with these other tools.

## 2.3  Requirements

Vannevar Bush envisioned the Memex to be a device that would augment the human memory (Bush, 1996). It is supposed to perform functions similar to that of a surrogate memory. We can say that the functional requirements of the Personal Memex will remain almost the same for the different populations we are considering. However, the user interface requirements of the device will differ for different populations. We discuss both these types of requirements in the following sub-sections.

### 2.3.1  Functional Requirements

Devices like the SenseCam may be used to *record* information. The main functions of the Memex in its interaction with the user may include – *recall*, *remind*, and *recommend*. In the background, the Memex would *organize information*.

- Recall – The Memex may help in recalling information. There should be facilities to perform the following:
  - o  Browse: Be able to explore or scan through information based on different organizations of recorded information.
  - o  Search: Be able to search over stored information using keywords or other metadata.
- Remind – The Memex may have a facility for providing reminders.
- Recommend - The Memex may provide recommendations following patterns in information that is recorded and then organized.
- Organize Information – The Memex would need to organize information based on associations made by the user explicitly (user inputs associations) or implicitly (through information gathered. For example, pictures captured by the SenseCam, coordinates captured by the GPS device, voice annotations at a particular instant of time.)

### 2.3.2  Interface Requirements

The Memex will need to *convey results* via a user interface. The design of the user interface will vary depending on the users of the device. From the literature review and the expert interviews, we classify user interface requirements into three categories – common user interface requirements, user interface requirements for those with macular degeneration and for those with MCI.

#### 2.3.2.1    Common user interface requirements

- Have sufficient guides/reminders to inform users where they are (in terms of the system) while using the device.
- Should help users be independent rather than completely dependent on the device.
- Monitor progress in tasks. Provide feedback to show progress in tasks.
- Capability to enable user to access/perform tasks quickly. For example, voice commands.

- Allow users to set and later change interface display attributes. For example, brightness, color, volume, etc.
- Make charging (and recharging) the device easy.
- Can the Memex help in recalling/remembering people, objects, events, etc in context (as they were situated/placed originally)?
- Compatibility with other assistive technologies that are currently available can be a big plus.
- Provide easy merging/integration with other devices/systems like calendar, email client, etc.
- The device should not lead to "learned helplessness" (refer Appendix that has interview notes).

### 2.3.2.2 Macular Degeneration

- Be able to adjust lighting conditions: People with macular degeneration prefer dim lit conditions than brightly lit conditions
- Contrast is important for them. They prefer high contrast.
- The design should provide for audio and visual input/output. There is a limit to the auditory memory. So the design should make sure that there's no overloading of any particular sense.
- There is no one template that can be universally used. We should provide for lot of customization. The menu design should be such that to access something there shouldn't be lot of steps involved. The items should be fairly easy to access.

### 2.3.2.3 MCI

- Avoid screen clutter or information overload on screens
- Have reminders to get back into focus on the task at hand
- Help in organization of information
- Some people have difficulty in recalling events based on time. Other cues such as people in those events, location of those events may help them recall events better. So, can the device help in recalling events based on other attributes apart from time?
- Make number steps to perform a task minimal
- It will be a good idea to audio reminders about things like taking medication. We should include the ability to go back and check (Did I take the medication?)
- For recall, it may be better to represent time graphically (like a timeline structure) versus numerically (like a table). Numbers are (anecdotally) problematic.

# 3   System Development

## 3.1  The System

The "Personal Memex" is a device that categorizes and organizes information for a user. It makes use of a SenseCam, which is a "badge-sized wearable camera' developed by Microsoft Corporation. The SenseCam will record, through the day, pictures, audio recordings and other sensor data. In addition to information from the SenseCam, the Personal Memex will also make use of other personal information like the user's calendar, phone, logs, email records and web browsing history. For our project, we assume that this massive amount of collected information is efficiently organized during the night and any part of it can be easily accessed. The interface to access information organized by Personal Memex can range from a cell phone, to a laptop PC, to a personal audio system, to a custom-built interface.

 Many applications come to mind for such a device, such as remembering people, alerting the user when she forgets to take a left turn on the road, providing details of a previously seen car, or reminding her where she left her keys.

For our project, we will focus on creating personalized view and recall function.

Our initial prototype has been developed for three population groups: those with mild cognitive impairment, those with macular degeneration and those without these disabilities.

## 3.2  The System Development Process

For this Personal Memex project, we considered a scenario based development framework (Mary Beth Rosson, 2002) which consists of analysis, design, prototype and evaluation. Although our system development process is based on scenario based development framework, some steps were modified to fit our project. Analysis echoes the phases of software development. In requirements analysis, the problem situation is studied through interviews with field experts, field studies of current, related systems and brainstorming. In addition, we considered findings from the literature review when developing the system requirements.

Envisioning new designs for the Personal Memex system began with a needs analysis to understand the problem. Based on our system objectives, we derived the general functions of the Personal Memex system. In the design phase, we developed some scenarios which provided a concrete glimpse of the system. They deliberately focused on functionality and user tasks. These scenarios were refined iteratively during the development process. In this project, we focused on a subset of functions, such as recalling a memory and creating a personalized view. All of the subsequent task and prototype development was limited to these two functions.

 The initial rough prototype was created with paper and pencil. Detailed system interaction was not specified in this initial stage. However, as the development and design phase continued, the prototype became more detailed and concrete. These prototype screens were evaluated  through cognitive walkthroughs by experts. Figure 3.1 shows the processes involved in the system development.

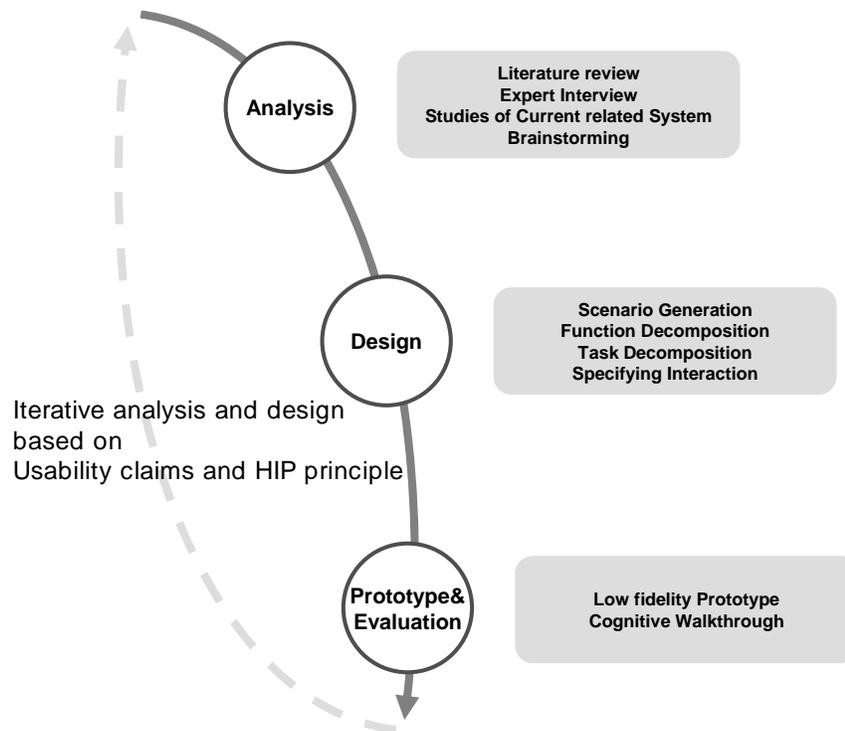

Figure 3.1 Overview of the system development process

## 3.3 Assumptions

The Personal Memex system will be based on the Sensecam wearable device which captures and records all data the user can see in some time interval. In addition we assume the following:

- The displays used for the Personal Memex system are based on PCs and/or PDAs. The prototype designed for this project is meant to be displayed on a PC or laptop display.
- Some AI technology such as voice recognition, computer reasoning, and expert systems will be applied to support the functionality of the Personal Memex system.

## 3.4 Scenarios

### 3.4.1 Population Notes

We considered both amateur athletes and musicians for our initial scenarios. There are common factors between athletes and musicians. They each have performances and coaches. Both are part of a team, yet they work on improving their individual skills. Athletes and musicians make an interesting user group for the SenseCam because their activities are not document-based in the manner of many office applications.

### 3.4.2   Personal Memex Functional Scenarios

These scenarios help to provide a quick overview of the system's overall functionality. Devices like the SenseCam may be used to *record* information. The main functions of the Personal Memex in its interaction with the user would include *recalling*, *reminding*, and *recommendations*. In the background, the Personal Memex *organizes information*. Finally, the Memex will *convey results* to the user via displays and other user interfaces (using vibrations and/or audio). Different sets of rules will be required to perform all these functions.

#### 3.4.2.1   Examples of Recall

Our initial scenario involves being able to replay a portion of a performance:
- A student practicing music can recall what his teacher told him about a particular section of music during his last lesson.
- An athlete can review a particular drill that his coach recommended.
- A musician can play along with the rest of the band, which makes practice more fun and realistic.  (Karaoke!)
- Coaches and band leaders can recall and review particularly well-done games or concerts.
- Sometimes the user may be involved in more than one task. Then, if the user needs to use all his/her attentional resources to focus on one task, the SenseCam may be used to keep track of the second task so that the user may pay attention to it later on. For example, a musician having to play his/her notes and also having to look at the conductor and other musicians at the same time.

#### 3.4.2.2   Examples of Reminders

- After reviewing past performances, remind the musician/athlete of the best time for creativity or to practice.  NOTE: Judgement of whether a performance was good or not would have to be done by annotation.  Ex. "Last Wednesday's game was especially good."  Or "Take the A-train played at last Saturday's gig went really well."
- The Personal Memex could analyze patterns that influence performance, such as diet, sleep, exercise.  (For example, "You win soccer games when you eat bananas.")
- Cues while performing or playing (For example, the words to a song.)  Another topic that might fall into this category is that of the signals that musicians send to each other. During a performance, musicians have signals for each other to know when to repeat, feature a soloist, or end the song.  Could the Personal Memex serve as a cue for similar messages for the MCI user?
- Reminders to eat before a game or a long meeting (For example, reminding runners to eat 4 hours before a race.)
- The Personal Memex could remind the user of which tasks are primary and which ones are secondary and help the user in switch back and forth and give attention appropriately.

**3.4.2.3    Examples of Recommendations**

The Personal Memex may provide recommendations from patterns discovered while the information is being gathered and organized.  This may include or require annotations from the user to determine which events, attributes or information are important.

- Which tasks do I perform well? When does my performance deteriorate?
- What is the limit on the number of tasks I can perform at the same time and the type of tasks I can perform?
- Which tasks are automatic for me and for which tasks do I need controlled attention?
- What is a good time of day (and place) for me to practice?
- What is a good time of day (and place) for me to create/compose?
- Sometimes, it is hard for us to focus on one task because there may be many signals that we need to respond to at the same time. In a task involving multiple sub-tasks, the Memex may help in figuring out which is the primary task and which the secondary.

**3.4.2.4    Examples of Organizing Information**

Task workload measurements could be used as metrics to measure the time and effort required for a task. To make things simpler, user annotations may be used to record these measurements. These measurements could then be used to organize information and for recall, remind and recommend functions.

For example, I have been giving musical performances for the last two months. How can the Memex help in organizing information so that I know:
- Musical pieces which I can play automatically
- Musical pieces on which I need to practice harder

## 3.4.3   Sample Scenarios

**3.4.3.1    Scenario Describing the Use of a Personal Memex by a Long Distance Runner**

April Greenside, a young woman in her late-twenties, has recently started running long distance races. April recently acquired a Personal Memex in the hope that it would help her in her daily activities and her new endeavor.

**3.4.3.2    Customizing/Personalizing the Interface**

April takes her new Personal Memex and decides to configure it to customize the interface of the Personal Memex. This will help her react/respond faster, better and more efficiently to the information displayed by the Personal Memex. Since she is more of an auditory kind of person and responds to auditory signals faster, she sets the alarm settings to a ring tone and an image to go along with it to remind her of the activity that she needs to perform (picture of a bus, when she needs to catch a bus, picture of a face, when she needs to meet a person). She also likes to associate locations to pictures of the locations, and sets the next level of detail to a map (how to get to that location).

(Other things to consider – codes for easy answer questions, colors for special kinds of alerts, a calming sleep music, for when she is trying to sleep, etc)

MCI: Instead of a ring tone, use a person's voice, or a beeping sound. Associate locations to people in those locations, rather than just pictures.

Macular Degeneration: In the pictures, instead of having too many details, zoom-in on the person/object in the picture. Use different bold colors for different reminders.

### 3.4.3.3    Starting a Query and Browsing

It has been over 4 months since April started using her Personal Memex. Today, she wants to use the Personal Memex to recall the instructions given by her doctor, two days ago. On her Personal Memex, she enters the specifications of her query terms as "Dr. Smith", "Tuesday", "Montgomery hospital" (could be pictorial, iconic drop downs). The Personal Memex brings up a series of pictures of her session with Dr. Smith. Using the links displayed by the Personal Memex, April, browses from the pictures of the session with Dr. Smith to pictures to a Yoga workshop (Dr. Smith had suggested some Yoga postures to strengthen April's back) that April attended two months back, to details about the instructor, the location, other people who attended the workshop with her, some of their phone numbers, etc. Thus, she moves along a path that the Personal Memex stores as "path visited".

MCI: Replay series of audio instruction files recorded by the Personal Memex and be able to browse from one audio conversation to another.

Macular Degeneration: Have bold, line diagrams of the Yogic postures instead of showing people. Have all the display bold.

## 3.5  Task Decomposition

### 3.5.1  Function Analysis based on scenarios

Based on the Personal Memex's primitive functions and the example scenarios (the narrative and highly abstract situational scenario), we analyzed the functions of Personal Memex and generated a detailed scenario.

First, we analyzed the Personal Memex system through the abstraction decomposition matrix derived from cognitive work analysis (Vicente, 1999). Abstraction-decomposition space, a two-dimensional modeling tool that can be used to conduct a work domain analysis in complex sociotechnical systems, is very useful for understanding the information requirements of systems and is shown below.

**Table 3-1 Abstraction-Decomposition Space**

| Goal-means | Whole-part | | |
|---|---|---|---|
| Purposes/constraints | Why ↑ | | |
| Abstract function | ↓ What | Why ↑ | |
| Generalized function | How | What ↓ | Why ↑ |
| Physical function | | How | What ↓ |
| Physical form | | | How |

The specific that apply to the design of the Personal Memex are shown below.

**Table 3-2 Phase 1: Description of System overview**

| Goal-means | Whole-part |
|---|---|
| Purposes/constraints | Helping the memory disparity |
| Abstract function | Reliability of the input memory, Stability, Time Criticality |
| Generalized function | Providing Information to User, Recording of User daily Memory, Organizing Information |

**Table 3-3 Phase 2: Decomposition of Whole part**

|  | System | Sub System | Component |
|---|---|---|---|
| Purposes/ constraints | Helping the memory disparity |  |  |
| Abstract function | Reliability of the input memory, Stability, Time Criticality |  |  |
| Generalized function |  | Providing Information to User, Recording of User daily Memory, Organizing Information |  |
| Physical function |  |  |  |
| Physical form |  |  |  |

**Table 3-4 Phase 3: Decomposition of Providing Information**

|  | System | Sub System | Component |
|---|---|---|---|
| Purposes/ constraints | Helping the memory disparity |  |  |
| Abstract function | Reliability of the input memory, Stability, Time Criticality |  |  |
| Generalized function |  | 3.0.0: Providing Information to User, 1.0.0: Recording of User daily Memory, 2.0.0: Organizing Information | 3.1.0: Recall*, 3.2.0: Remind**, 3.3.0: Recommend*** |
| Physical function |  |  |  |
| Physical form |  |  |  |

*Recall: Providing the information or memory related to time dependant skill and knowledge to support intended task
**Remind: Providing information or memory related to event dependant skill and knowledge
***Recommend: Helping to select the task alternative

**Table 3-5 Phase 4: Decomposition of Recall**

| | System | Sub System | Component |
|---|---|---|---|
| Purposes/ constraints | .. | | |
| Abstract function | .. | | |
| Generalized function | | Providing Information to User, .. | Recall*  Remind**, Recommend*** |
| Physical function | | | |
| Physical form | | | |

**Table 3-6 Phase 5: Focused on two subfuctions**

| | System | Sub System | Component |
|---|---|---|---|
| Purposes/ constraints | Helping the memory disparity | | |
| Abstract function | Reliability of the input memory, Stability, Time Criticality | | |
| Generalized function | | 1.0.0: Recording of User daily Memory, 2.0.0: Organizing Information 3.0.0: Providing Information to User, | 2.1.0: Automated Organizing 2.2.0: **Creating personalized view** 3.1.0: **Recall,** 3.2.0: Remind, 3.3.0: Recommend |
| Physical function | | | |
| Physical form | | | |

### 3.5.2   Example for Specification from Function to Task

#### 3.5.2.1     Judgment of the task situation

Implication: User of the Personal Memex needs to understand the task situation.

Issues and assumptions:
- If the Personal Memex judges the situation, how can the Personal Memex be aware of the situation?
  : Personal Memex judges the situation using user environment data (place, time, user's daily schedule and etc.) in the example of long distance runner, Personal Memex can judge the situation which user will go running after breakfast using user's email.

- If the User judges the situation, how can user recall the situation-related memory?
  : User can make a setting of system. For example; if the user has a serious memory disparity, the user can create a setting in which the system automatically provides a recollection about all situations. If user has a minor memory disparity, the user can make a setting which system returns a recollection according to the user's specific request.

- When the user is aware of the task situation, how can the user input the situationally related information to the Personal Memex?
  : User can input the situation related information using user voice input or writing on the system screen.

- When the user is aware of the task situation, what kind of information should the user input into the Personal Memex system?
  : User has to input the information necessary to elicit the data related task.

#### 3.5.2.2     Elicitation of the memory

Implication: the Personal Memex instantly retrieves and displays useful, task-related information from the semantic network of organized information using the organizing function (Function 2.0.0 shown on Table 3-6).

Issues and assumptions
- If the Personal Memex retrieves the information from the semantic network, how can the Personal Memex determine the priorities of this related information?
  : The priorities of the information is determined by the user or by system algorithms.

#### 3.5.2.3     Providing the information

Implication: the Personal Memex requires a method to provide the user with situation-related information.

Issues and assumptions

- What kind of alternative methods are available for the Personal Memex to present the information?

: the Personal Memex can use various methods such as vibrations, haptic or tactile outputs, electronic signals, audio playback, and visual displays.

### 3.5.3 Task list of Personal Memex based on prototypical process

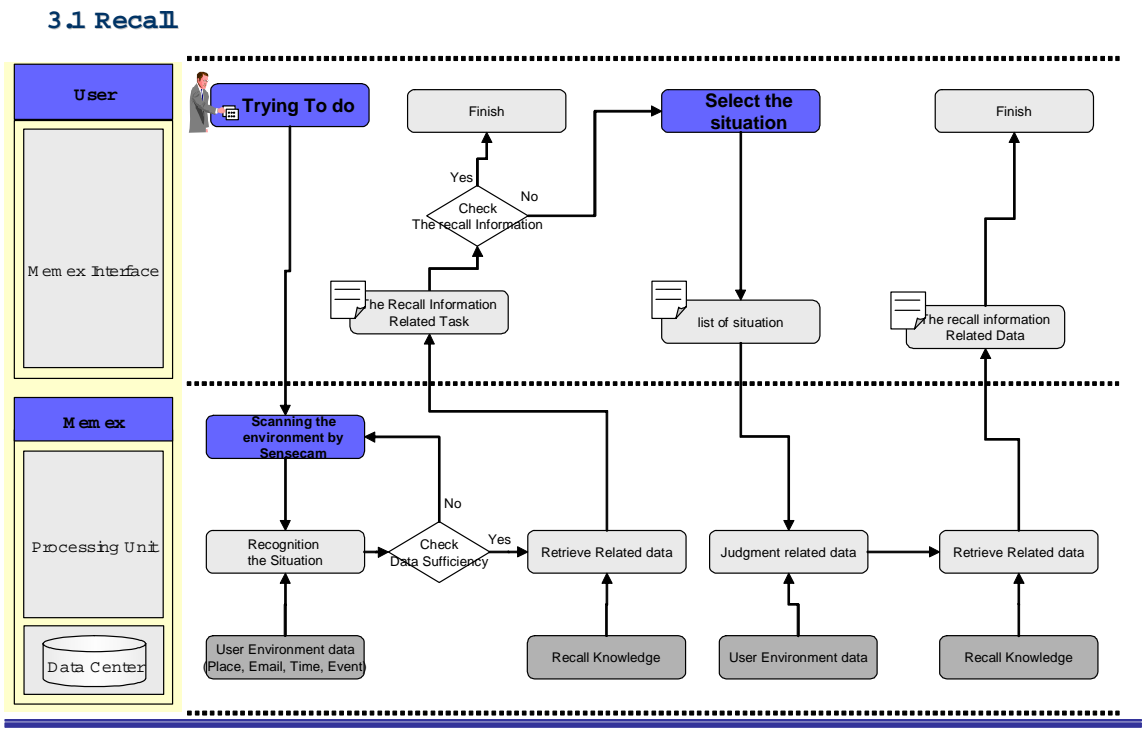

**Figure 3-1 Protypical Process Chart**

## 3.6 Prototype Design

Based on the task decomposition, the parts of our prototype can be envisioned via the following concept map.

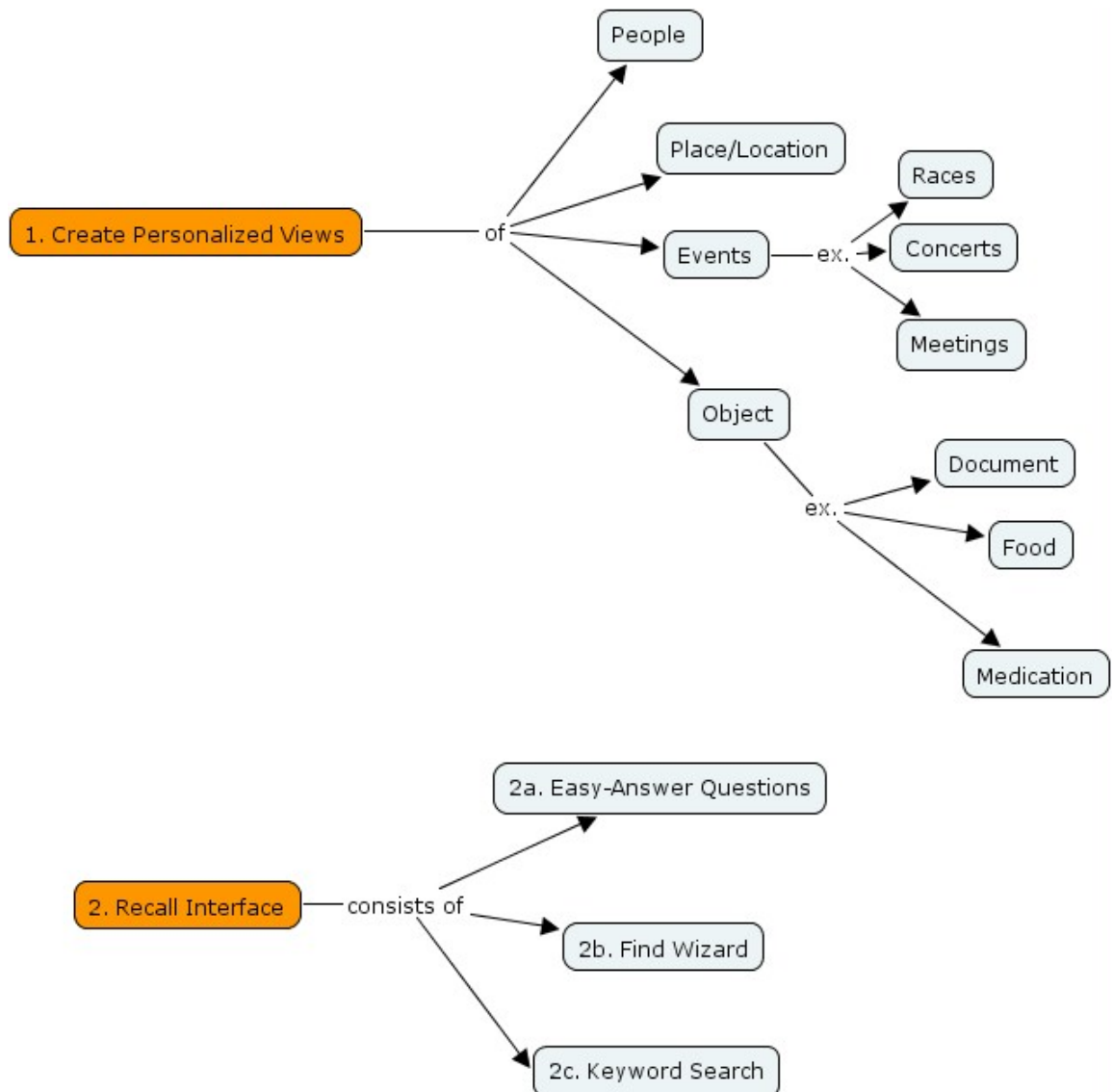

**Figure 3-2 Concept Map of the Prototype's Functions**

Screens were drawn for each step and are shown below. Individual screens will be discussed in more detail in section 4 where the final prototype, which is very similar to the prototype used for the walkthroughs, is shown.

### 3.6.1 Customization Screens

The first set of screens allowed a caregiver or the user to customize the Personal Memex.

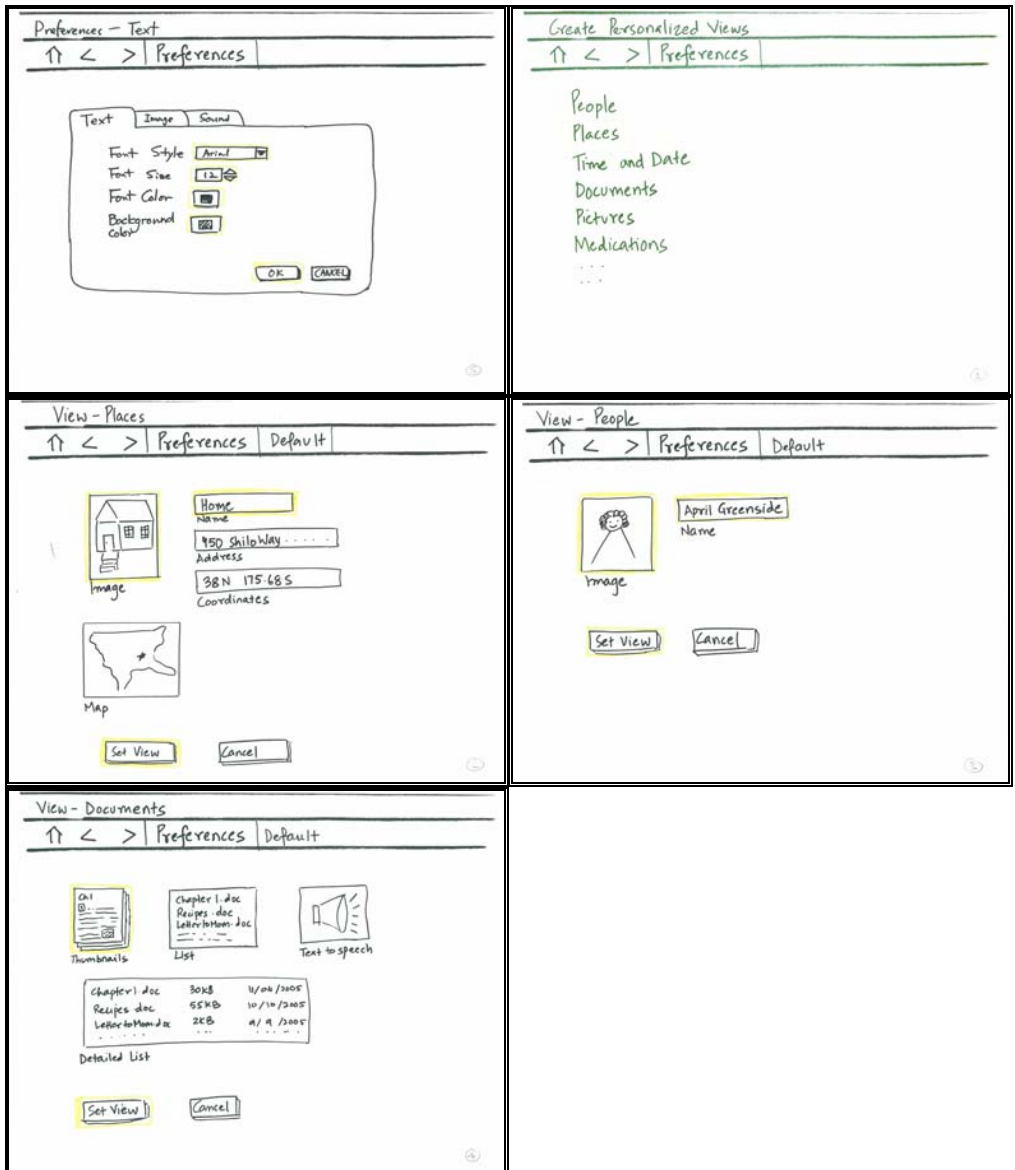

**Figure 3-3 Initial Preference Screens**

## 3.6.2 Find Screens

The second set of screens showed how April, our runner, could recall details from a conversation with her doctor.

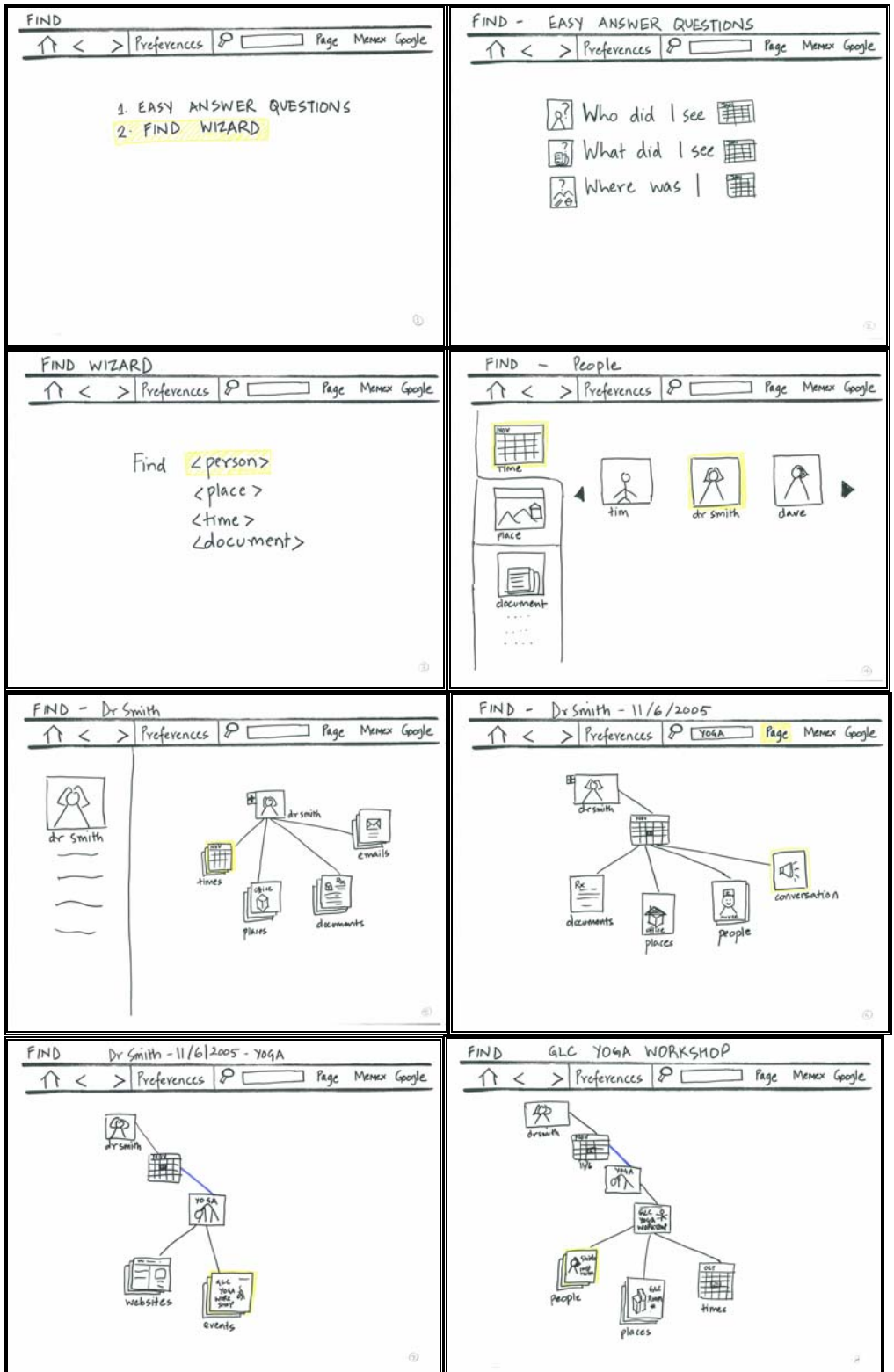

**Figure 3-4 Initial Find Screens**

## 3.7 Walkthroughs

### 3.7.1 Cognitive Walkthrough Process

A Cognitive Walkthrough consists of the following four steps:

1. Convene the analysts and Create the Scenarios
2. Perform the walkthrough
3. Record the usability problems found
4. Fix the usability problems

Our approach is shown in Figure 3-2.

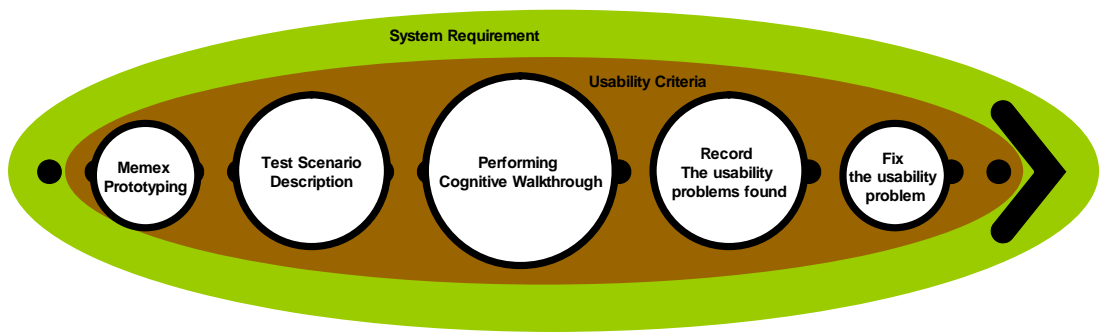

Figure 3.2 Cognitive Walkthrough Process

### 3.7.2 The Analysts

Karen Roberto
Dr. Roberto is a Professor in the Department of Human Development and Director of the Center for Gerontology. Dr. Roberto's research focuses on health and social support in the later stages of life.

Bill Holbach
Assistive Technologies Coordinator

Hal Brackett
Assistive Technologies Lab and Special Services Manager

Virginia Reilly, ADA Coordinator

### 3.7.3 The Scenario Description for Cognitive walkthrough

We prepared the following scenario for the cognitive walkthrough, gave it to the analysts and also explained it as we walked through each part of the prototype.

**Overview**

*Who*

April Greenside, a young woman in her late-twenties, has recently started running. She visits Dr. Smith, who gives her advice on exercises and diet, and suggests that she try yoga. Dr. Smith feels that yoga will help her become more flexible and strong. April recently acquired a Personal Memex in the hope that it would help her in her new endeavor.

*What and Why*

April Greenside has decided to use Personal Memex to retrieve the information related to Dr. Smith' suggestions for flexibility and strength. The Personal Memex has recorded Dr. Smith's instructions and has organized it. April wants to find the simplest way to use the Personal Memex to recall Dr. Smith's instructions.

*With*

Personal Memex

*How*

To query the required knowledge (Dr. Smith's recommendation), April uses the "Find Wizard" option on the main screen. She pulls out the stylus pen, clicks keyboard icon and taps the alphabet key to input her question. And then, she clicks the search button.

April has the option to use an Easy Answer Question to recall Dr. Smith's instructions. (This option is not shown in the prototype).

April Greenside can input the situation through speech. She says what she wants to do. And the Personal Memex parses her speech and give proper feedback.

April can also create personalized views and preferences.

### 3.7.4  Task Description

We gave these task descriptions to our analysts to show them how the Personal Memex could be used in the scenario.

Recall Task

1. Find the instructions April received from Dr. Smith from her last visit.
2. Review yoga postures and exercises.
3. Find some information about yoga resources in the local area.  (Are there   classes, instructors, etc?)

Creating Personalized View

1. Select *image* and *name* in personalized view of places.
2. Select *image* and *name* in personalized view of people.
3. Select *thumbnail* in personalized view of document.
4. Using Preferences, make the document text size bigger, highlight text, and change the background color.

### 3.7.5  Guided Question for Cognitive Walkthroughs

After going through the prototype with the tasks listed above, we gave the analysts these questions.

***Examine the sub-goals***

1) Will the user try to achieve the right sub-goal?
2) What knowledge is needed to achieve the right sub-goal?
   Will the user have this knowledge?

***Examine the actions needed to satisfy these sub-goals***

3) Will the user notice that the correct action is available?
4) Will the user associate the correct action with the sub-goal she is trying to achieve?

***If the correct action is performed, examine the feedback and the goal hierarchy***

5) Will the user perceive the feedback?
6) Will the user understand the feedback?
7) Will the user see that progress is being made towards solution of her task in relation to her main goal and current sub-goals?
8) Is it obvious what the system is trying to do?

### 3.7.6  Cognitive Walkthrough Interview Notes

As the walkthrough progress, the members of the group recorded the audio portion of the walkthrough and took notes.  The notes are summarized in Table 3-7 below.

**Table 3-7 Notes from the Walkthroughs**

| Feature | Comment | Solution / suggestion |
|---|---|---|
| Preferences | Could get stuck if font and background are accidentally set to be the same | Put in two graphic areas showing the 'Current' settings and then another for the 'New' settings feedback* |
| Preferences | Resetting to Default is not available at top level | Add Reset button * |
| Preferences | Highlighting requires contrast | Add better visual / audio feedback + |
| Preferences | Vague icon for text to speech | needs to be clarified. For example, if you are looking at a list of files, does it read the list?* |
| Audio | Preferences for speech frequency, speed | Include in screen for audio preferences*,+ |
| Audio | Audio summary of piles vs. full reading of full document | Include in screen for audio preferences* |
| Icons at top (fwd/bk/home) | Requires familiarity/ training | Tooltips, Fish-eye (eg hyperbolic magnification) hover (mouse-over) * Add text labels + |
| Terminology | Too computerese, generational | Default => 'Standard' Documents => 'Papers' Specialists => 'Doctors' + |
| Interactive Calendar | What is today? | Always Highlight 'today' – current date and time* |
| Easy Answer Qs | When did I see John last? | Add combination questions like this* |
| Easy Answer Qs | When did I last take xyz Medication? | + |
| Find Wizard: Surfing associations | Good feature | *,+ |
| Find Wizard: searching & sorting by context | Context & associations are key for recall What about searching audio? | make sure that status is visible depending on where the search is being executed : ie page or pile * Make the current search context (scope) visible (highlight it) + |
| Links between piles | | Change color if it was already searched or visited* |
| Minimizing / collapsing a pile | Need to reduce visual clutter / overload | Add [-] button at each pile's uplink to correspong with [+], use fade/transition + |
| Picking up trails | Might want to save or bookmark a search state | Add a 'bookmark' feature |
| Searching | What if the target is not found? | Consider fallback screens + |
| Date | To cue people about which day is 'today' and which is 'yesterday.' | highlighting today's date* |
| Date | Can I scroll forward to other dates in the future so that I can see my future events? | * |
| Input | Mark Important Events (in time) | Need simple, natural input* |
| Input | Add something to schedule, TO DO reminder | Need simple, natural input+ |
| Compatibility | With other local machines, might want to print | Add print button, Use Bluetooth, * |
| Using the device | Already remembering cell phone, PDA, Watch, emergency pager | RFID and alert w/o proximity to glasses, watch* Consider medication awareness + |
| Help | None is available | Add, but do not loose state* |
| Tutorial | Need one for Rehearsal, 'monkey see, monkey do' | Include Examples of use* |
| Text and Icon | suggested mouse-over magnification of text | * |

+ Roberto
* Reilly, Holback, Brackett

# 4 Results

## 4.1 Final Prototype

In this section, we present the final prototype along with a description of each screen. Comments and suggestions from experts have also been included. In the "General Comments from Walkthrough" sub-section, we discuss comments and suggestions by experts that we have not shown in the final prototype.

### 4.1.1 Create Personalized View

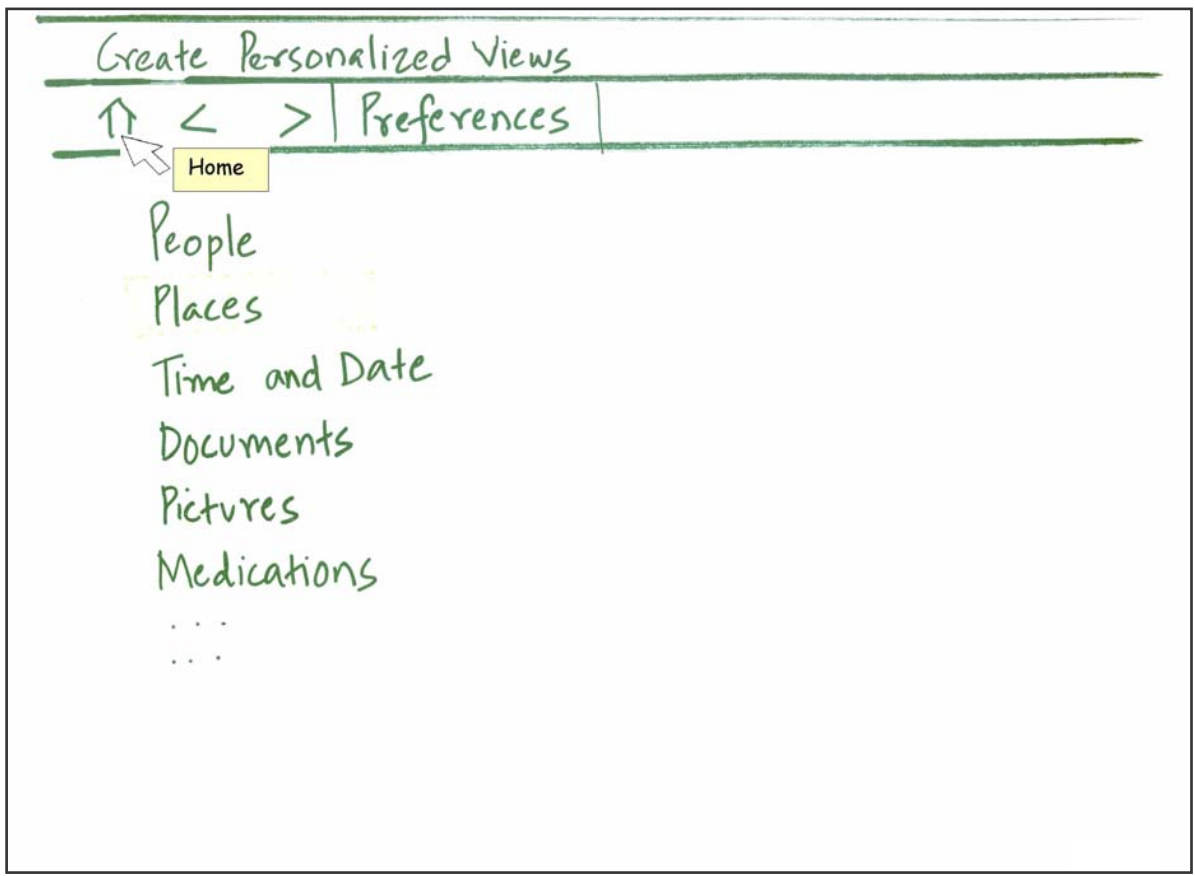

**Figure 4-1 Create Personalized View**

This functionality lets the user create personalized views of different data types. For example, some people associate people with their pictures or a combination of their picture with their name. The "Create Personalized View" allows the user to make such settings. In the example shown in the screen (Figure 4-1), the user selects to create a personalized view of "Places"..

Our initial prototype had icons for screen navigation – "Home", "Back", and "Forward". In the walkthrough, the experts suggested that we add text descriptions to these icons since some populations understand words better than symbols. To reduce screen clutter, we added tooltips to describe these icons.

### 4.1.2   View – Places

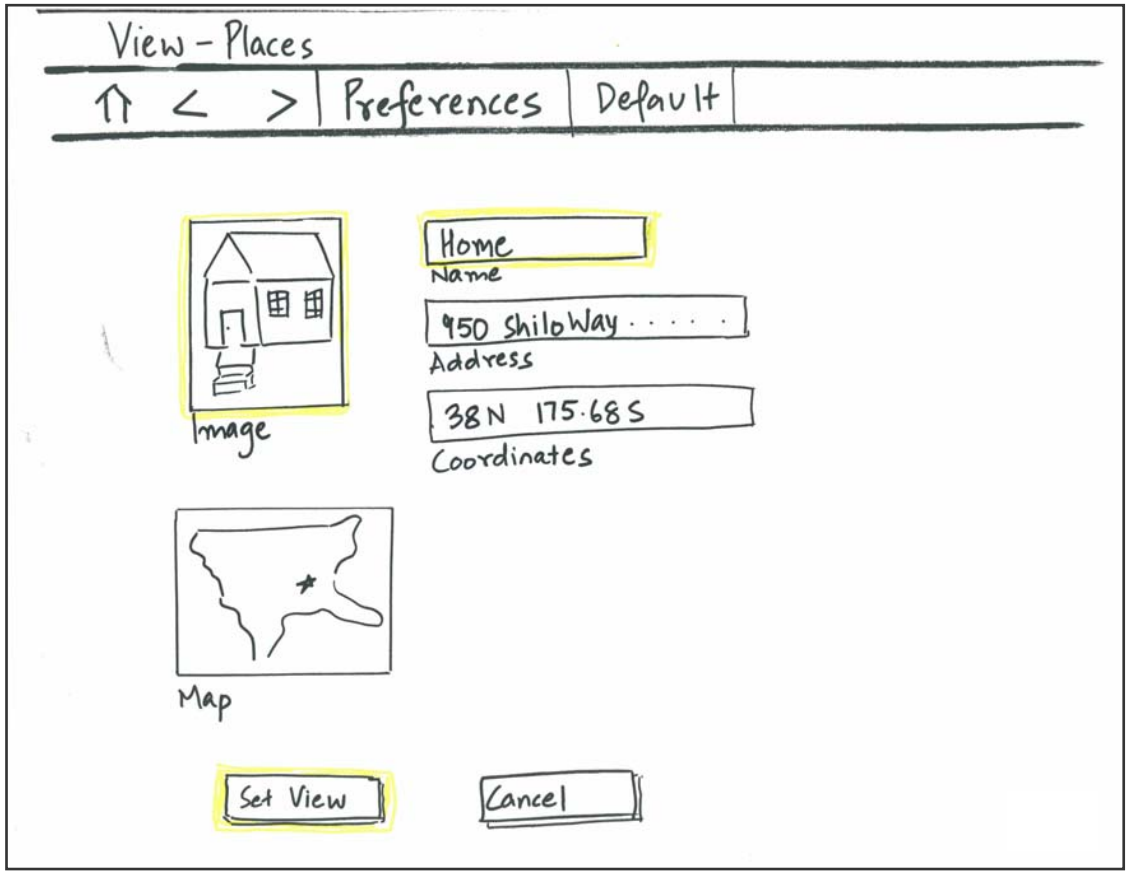

**Figure 4-2 View – Places**

Views for "Places" include image, name, address, coordinates and map. The user may select one or a combination of more than one view for "Places". In the example in the screen (Figure 4-2), the user selects the combination of image and name and then hits on the "Set View" button.

The "Default" option in the toolbar (for all View screens except the main "Create Personalized View" screen), allows the user to set a default view for a particular type. In our prototype, we consider the default view to be the manner in which that type was recorded. For example, the default view of a conversation may be audio, for a place it may be an image, for a document it may be a thumbnail.

### 4.1.3   View – People

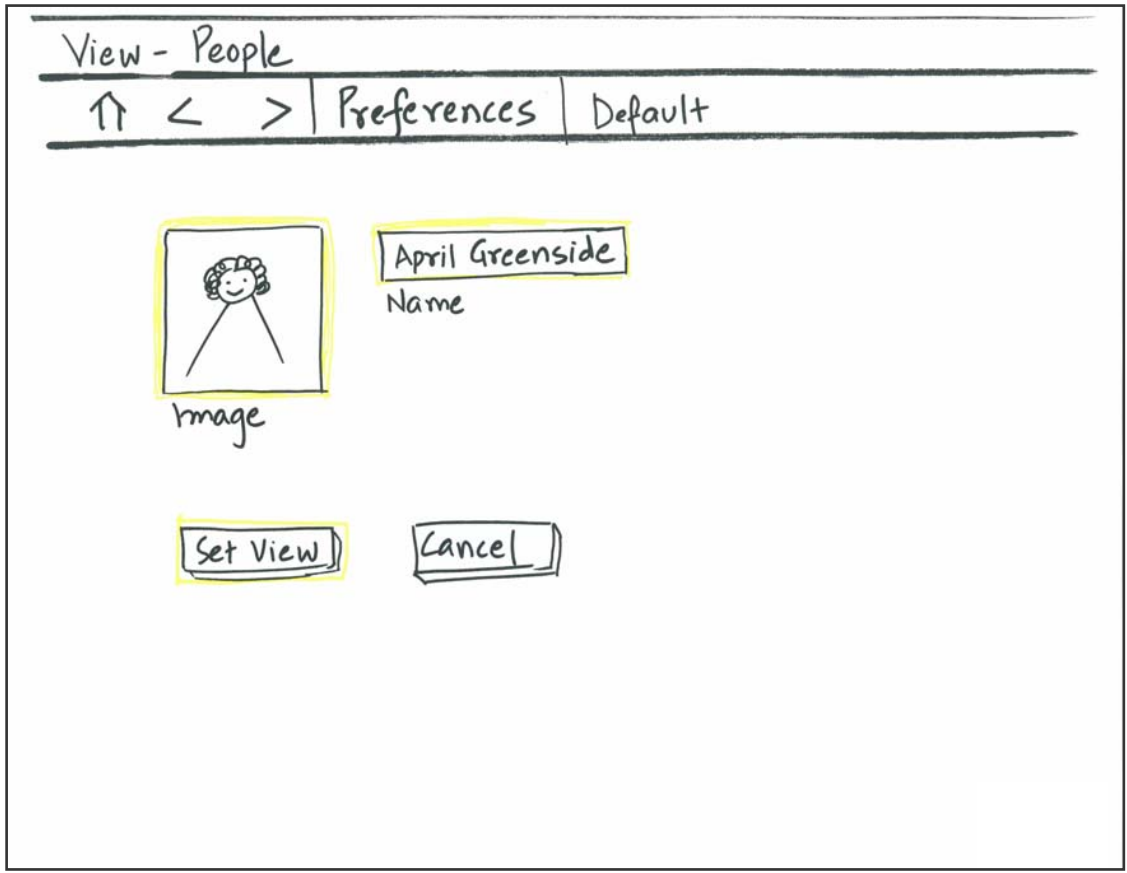

**Figure 4-3 View – People**

Views for "People" include image and name. In the example in the screen (Figure 4-3), the user selects the combination of image and name.

### 4.1.4 View – Documents

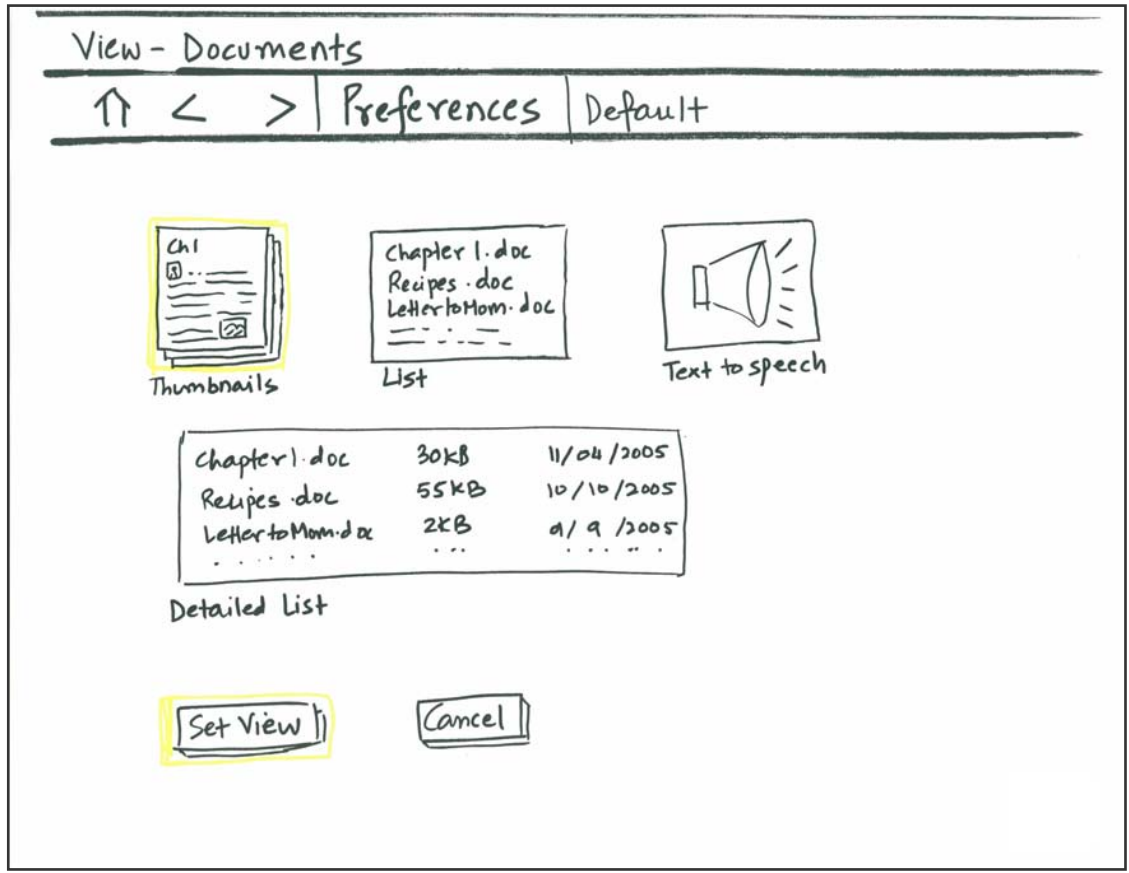

**Figure 4-4 View-Documents**

Views for "Documents" include thumbnails, list, detailed list and text-to-speech. In the example in the screen (Figure 4-4), the user selects thumbnails.

The text-to-speech option in this case may provide an audio summary of lists of documents. Individual documents may also have a text-to-speech option, where the contents of the document may be converted to audio. This feature will help users with visual impairments.

### *4.1.5   Preferences – Text*

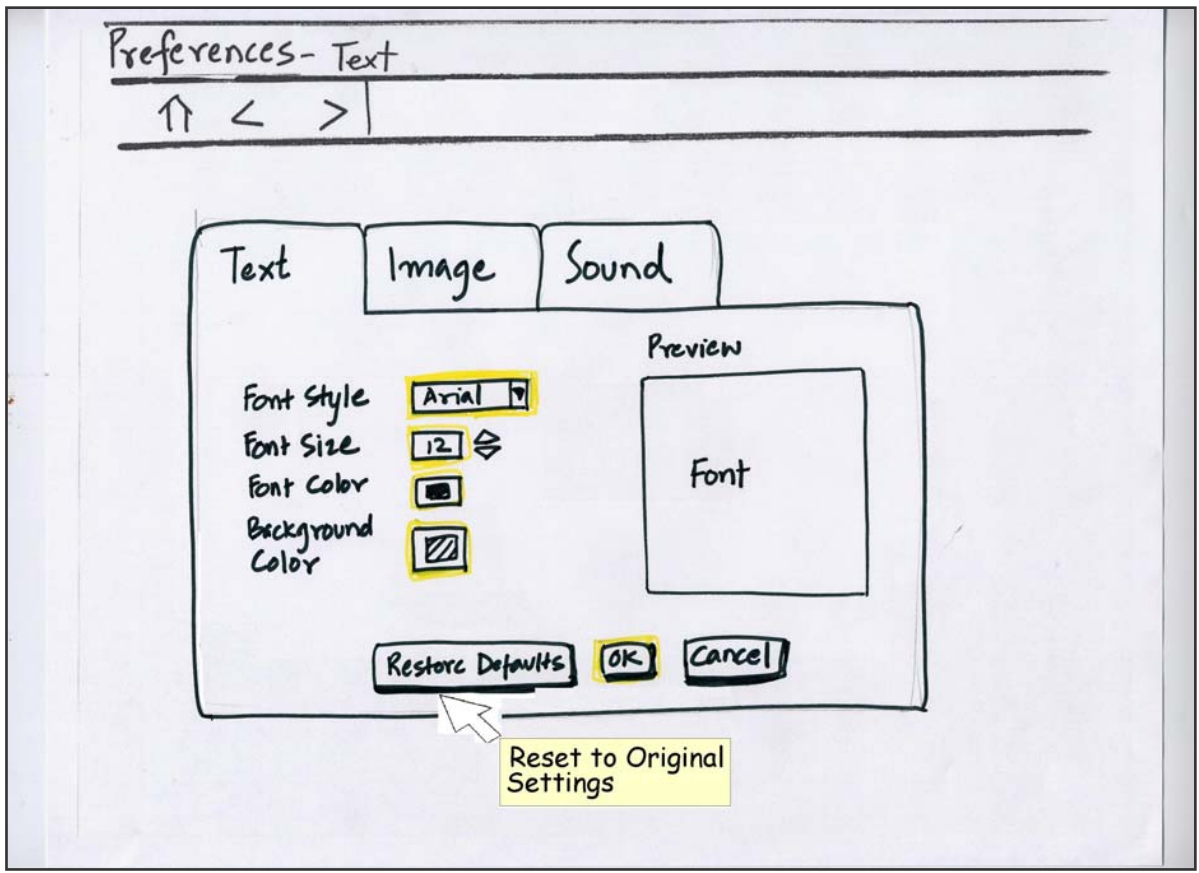

**Figure 4-5 Preferences – Text**

This screen lets the user set preferences for each data stream type. The example shown is that of text preferences. Sound preferences could include setting the volume, tone, frequency, etc. Image preferences could include contrast, color or black/white, brightness, etc.

The "Preview" pane and the "Restore Defaults" button were added after the cognitive walkthrough. The "Preview" pane shows the preview of the settings made by the user. It provides feedback based on the choice of settings made by the user. The "Restore Defaults" button resets preferences of a particular data stream to its original settings. This feature will allow the user to rollback to original settings incase the user accidentally makes a wrong choice (for example, black text on black background).

### 4.1.6   Find

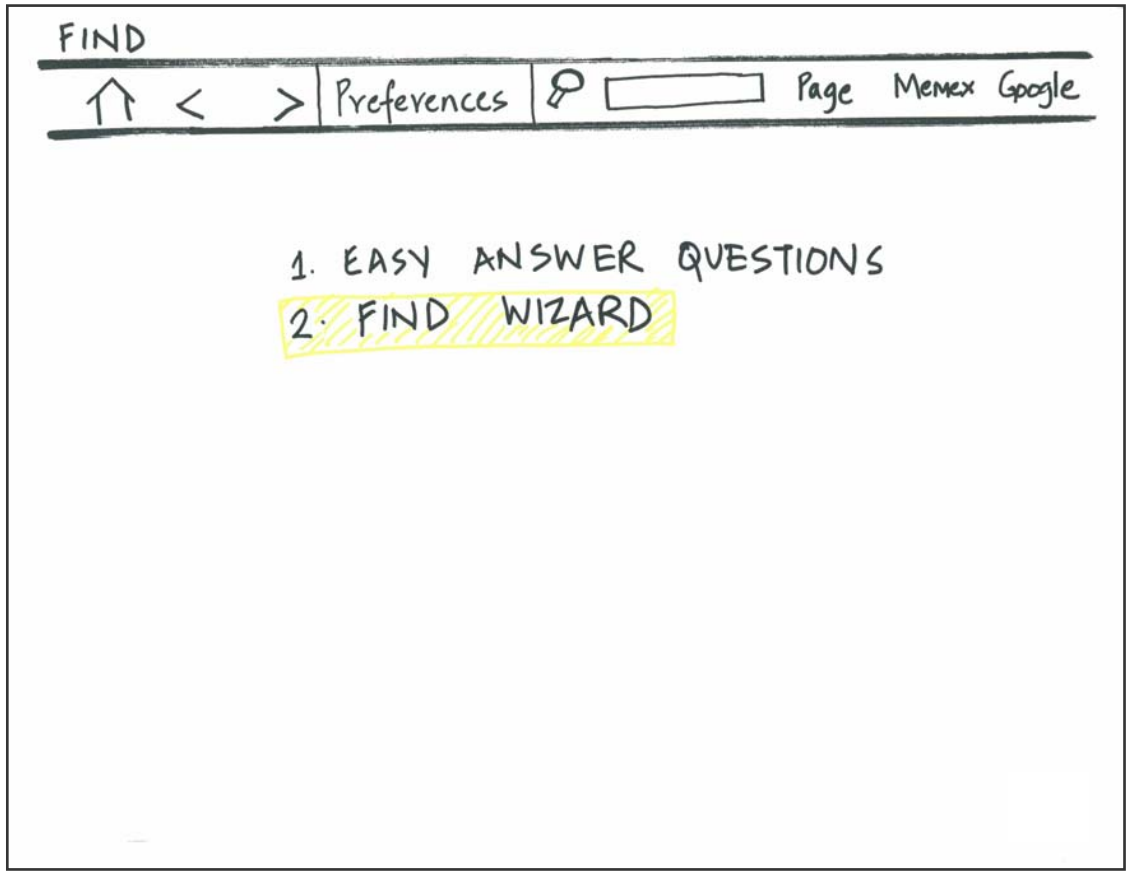

**Figure 4-6 Find**

The "Find" screens enable the user to recall previous information. Each "Find" screen has the home, back, forward, and Preferences buttons. In addition, there is a search toolbar, which can be used to search the contents on the current page, the Memex and Google (intended to be the world outside the Personal Memex).

The "Find" screen has two options, "Easy Answer Questions" and the "Find Wizard". The user in the scenario, April, selects the "Find Wizard" option.

### *4.1.7 Find – Easy Answer Questions*

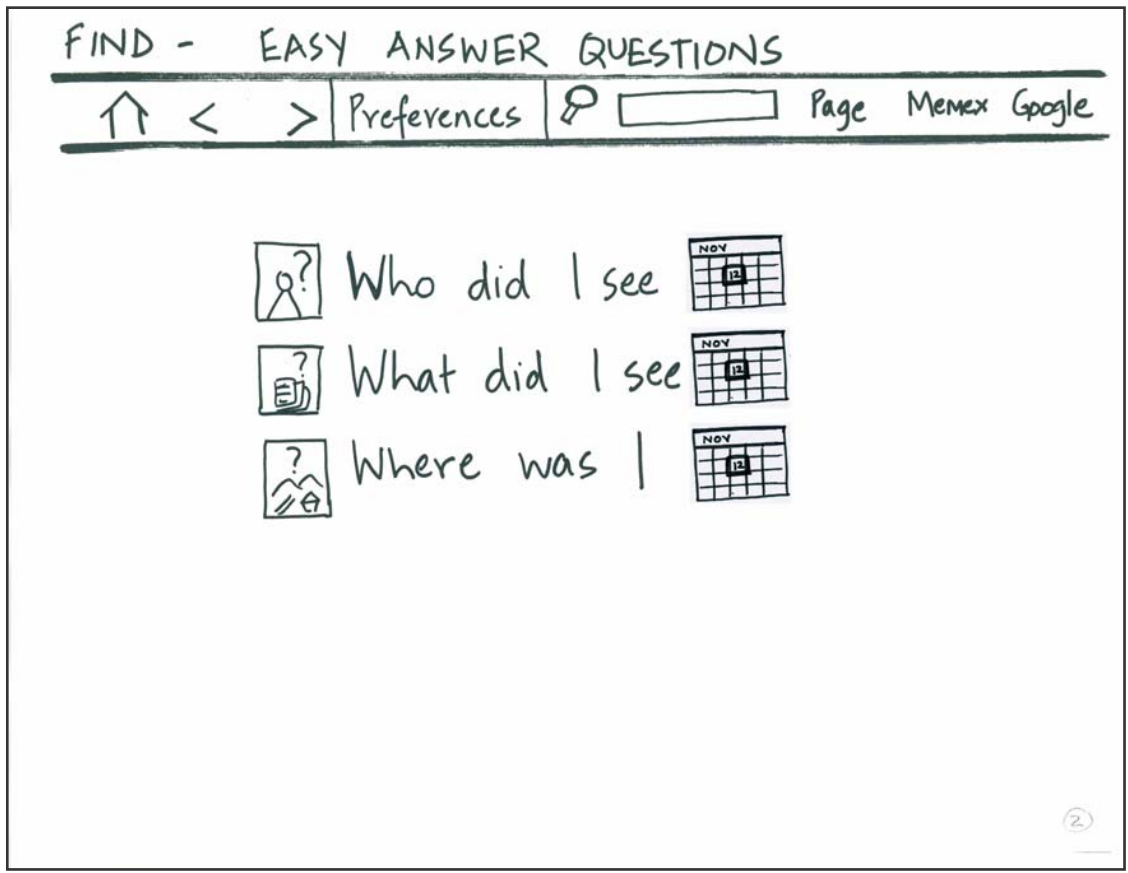

**Figure 4-7 Find - Easy Answer Questions**

Easy answer questions will help the user recall quickly, information that she needs often. Examples of easy answer questions include "Who did I see", "What did I see", etc. The current date is highlighted to cue the user and provide a time reference about which date is "today" (and hence, "yesterday", "day-before-yesterday", etc).

### 4.1.8 Find Wizard

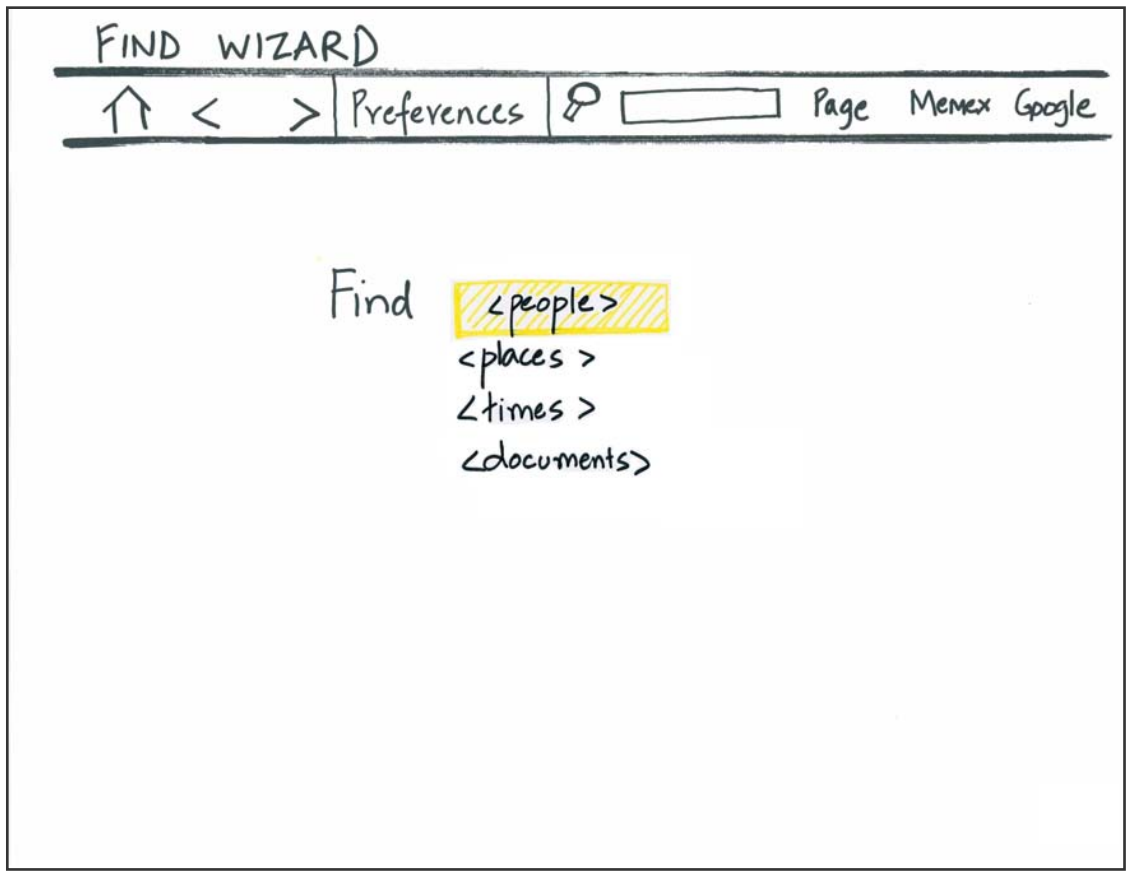

**Figure 4-8 Find Wizard**

The Find Wizard feature enables the user to recall information using a browse-like service. As the user goes through the Find Wizard to recall information, she may narrow down the information space over which she browses by selecting from different options. With the help of the Find Wizard, the user may build a trail of associated information

In the "Find Wizard" screen (Figure 4-8), the user may choose from different data type options like people, places, times, documents, etc. In our scenario, April wants to recall information from a meeting with her doctor and she selects "People" on this screen.

### 4.1.9   Find – People

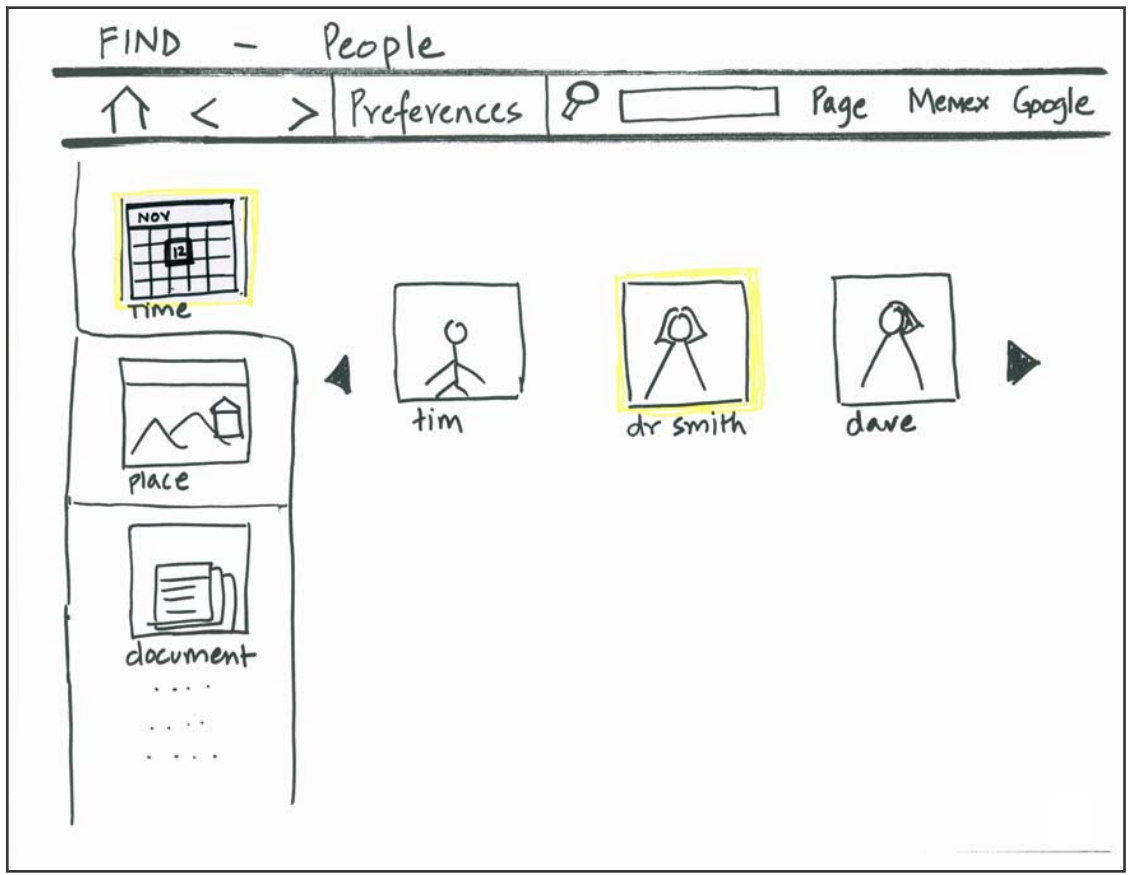

**Figure 4-9 Find – People**

The Find – People screen (Figure 4-9) consists of a scrollable list of People that April met, ordered by time. The list can also be ordered by places and document and other types by choosing the appropriate tab.

April finds and selects Dr. Smith, her doctor, to get all information in the Personal Memex related to Dr. Smith.

### 4.1.10 Find – Dr Smith

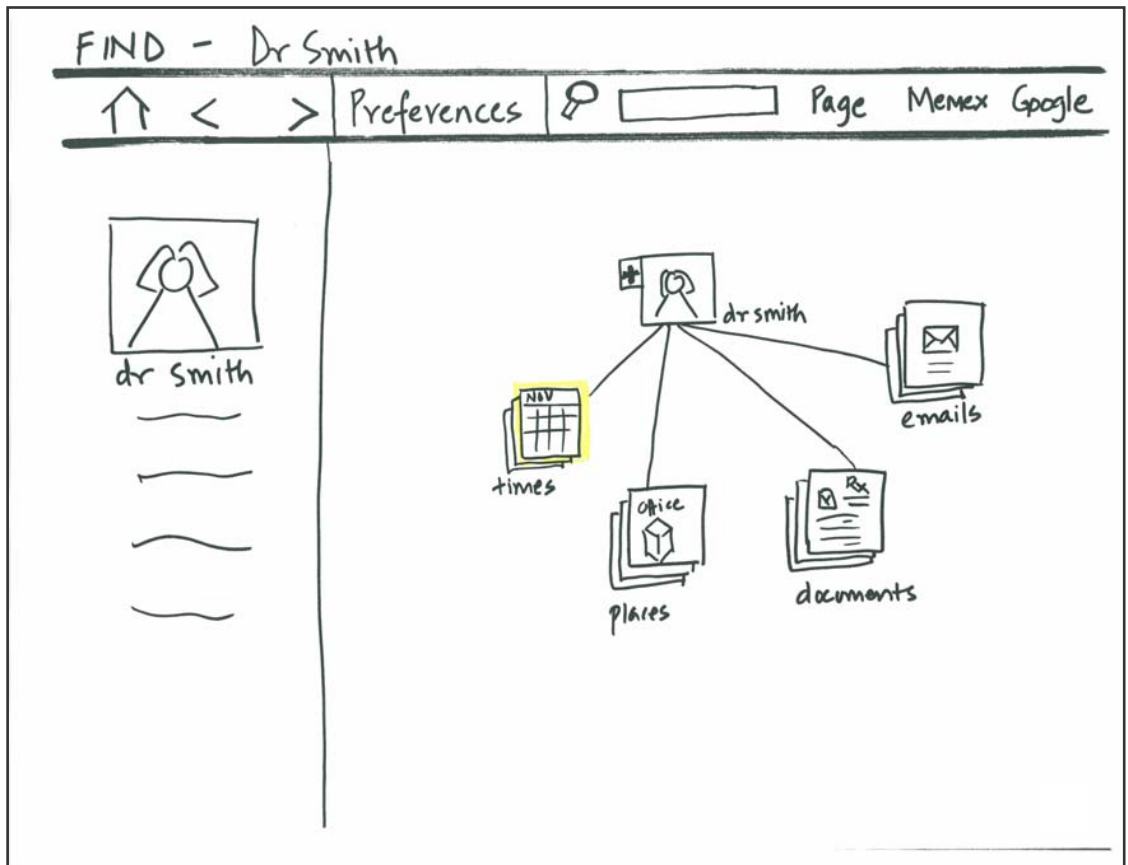

**Figure 4-10 Find - Dr. Smith**

The "Find – Dr. Smith" screen (Figure 4-10) shows details of Dr. Smith along with a tree-like structure that shows all information associated with Dr. Smith in piles based on the type. April chooses the latest meeting with Dr Smith, on November 6, from the "Times" pile.

The "+"sign next to the image representing Dr Smith in the tree indicates that there is more information associated with Dr Smith. The sign may be selected to view this information. This feature help avoid screen clutter.

### *4.1.11 Find – Dr Smith – 11/06/2005*

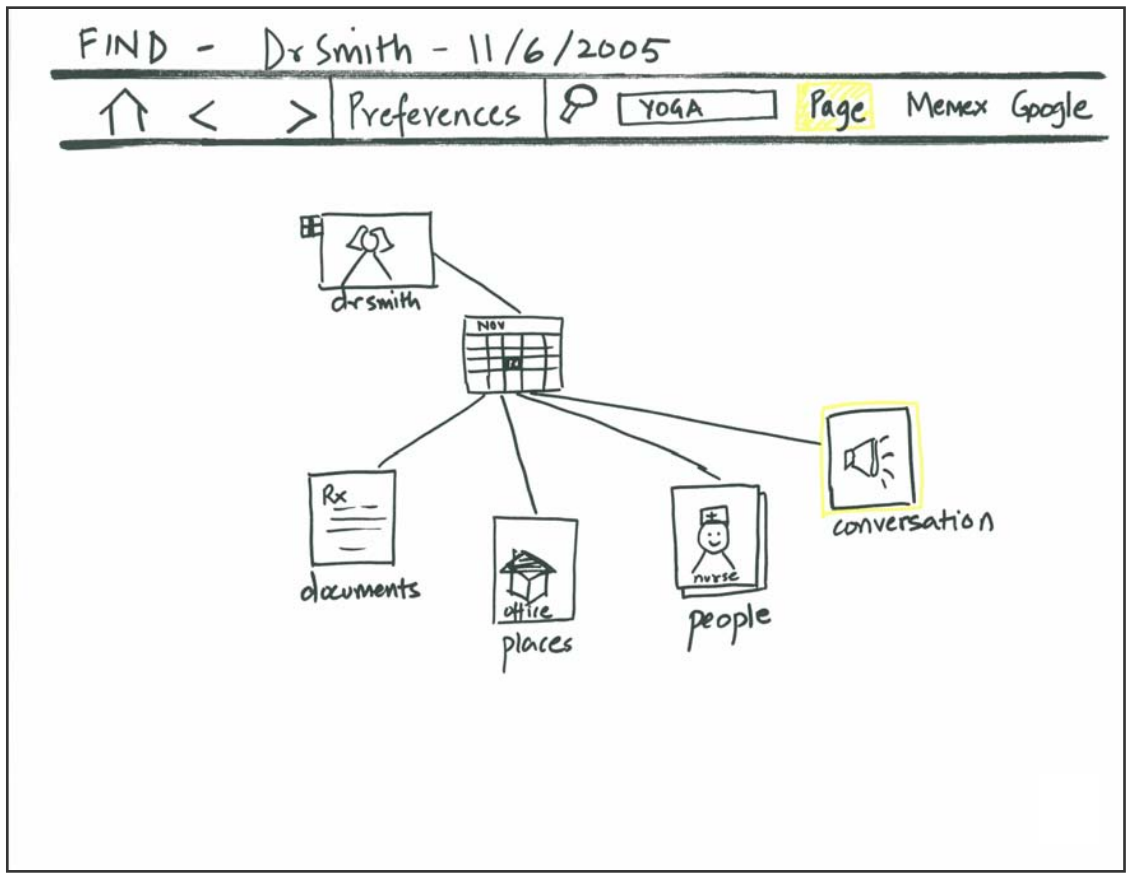

**Figure 4-11 Find - Dr. Smith - 11/6/2005**

This screen (Figure 4-11) shows all information associated with April's meeting with Dr. Smith on 11/06/2005. It also contains the items that April selected to follow this trail – in this case, Dr. Smith.

April wants to recall some instructions that Dr. Smith gave her about Yoga exercises. She uses the search box to search the conversation for Yoga related instructions. By selecting the "Page" option in search and conversation item, she restricts the search context to the current page in the Find Wizard.

### 4.1.12 Find – Dr Smith – 11/06/2005 – Yoga

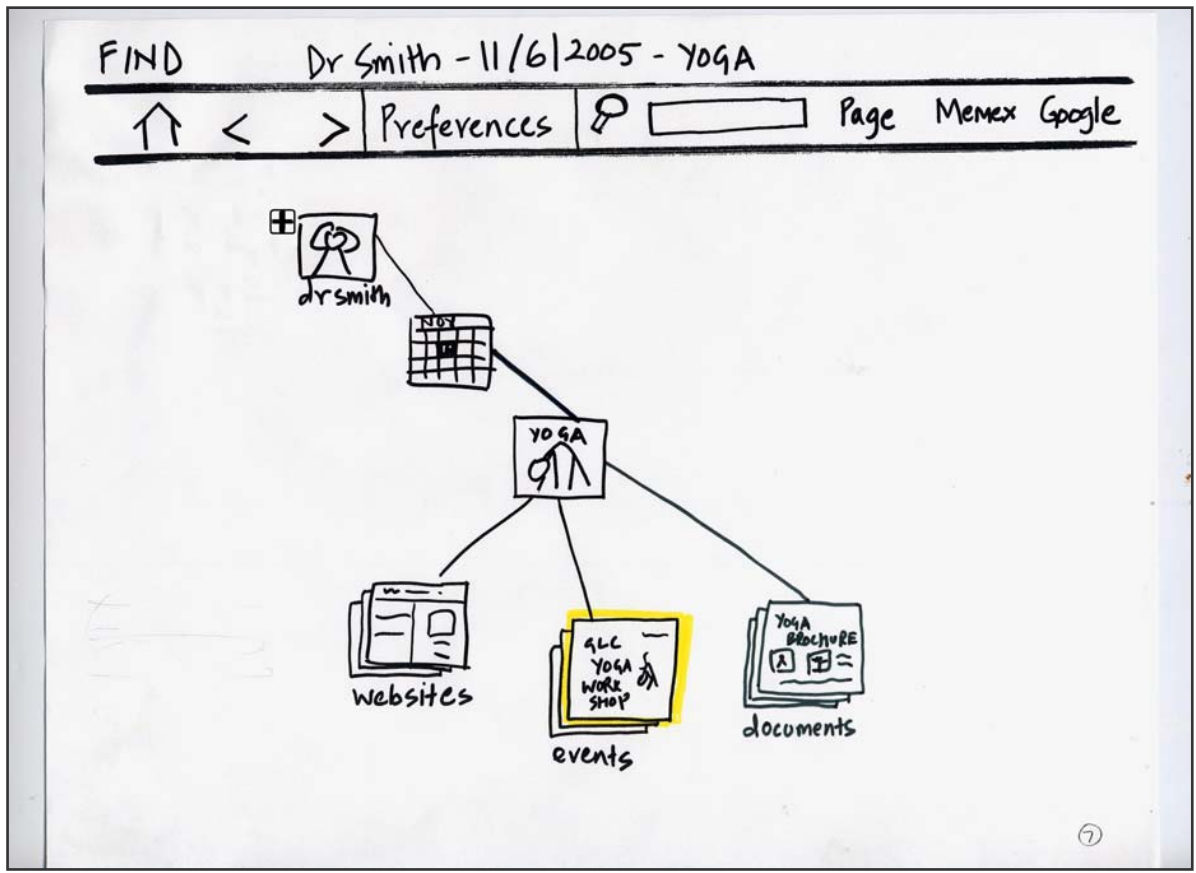

**Figure 4-12 Find - Dr. Smith - 11/6/2005 – Yoga**

This screen (Figure 4-12) shows the Yoga instructions within April's conversation with Dr. Smith on 11/06/2005. It also shows all information associated with Yoga in April's Personal Memex.

In our initial prototype, we indicated trails that were obtained by search (versus browse) using a different color. Based on comments and suggestions from experts, we removed this feature and indicated all trails in the same color. Experts felt that this information may be unnecessary to the user in her tasks and may confuse her. As per the scenario, April wants to recall information on the GLC Yoga workshop that she attended some months ago. She chooses the same from the "Events" pile.

### 4.1.13 Find – GLC Yoga Workshop

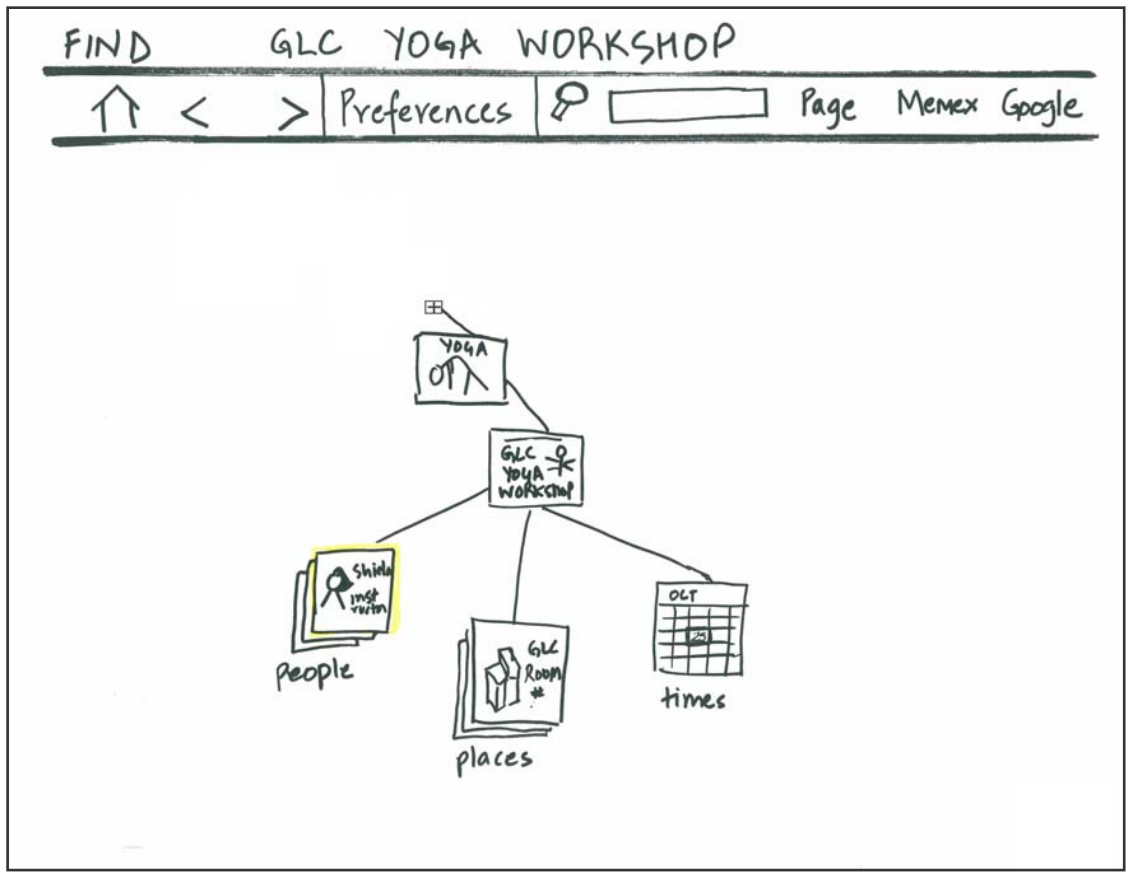

**Figure 4-13 Find - GLC Yoga Workshop**

The "GLC Yoga Workshop" screen (Figure 4-13) has details about the workshop. The "+" sign above the Yoga icon indicates that the trail expands further. This feature helps to avoid screen clutter. April wants more information on Sheila, the Yoga workshop instructor, and chooses her from the People pile.

### 4.1.14 General Comments from the Cognitive Walkthrough

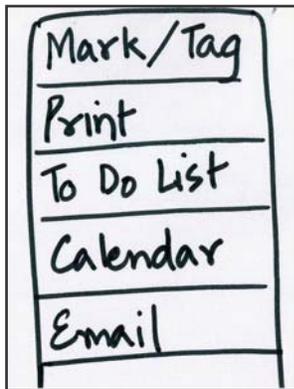

**Figure 4-14 Context Menu**

One suggestion for the "Find" screens was to link those screens to other applications and devices like the user's calendar, to-do list, email client, printer, etc. For example, when the user sees information associated with the "GLC Yoga Workshop", she may want to add any future event at GLC to her calendar, or she may want to email details of a forthcoming event to a friend. Another suggestion was to allow some kind of marking or tagging of trails. In our prototype, we assume that all trails formed by the user will be saved. However, if the user wanted to mark a trail as "Important", "Used Frequently", "Include in Easy Answer Questions", etc, there should be some facility to do so. As a workaround, we decided to add these features in the menu. Using the menu (not shown in the prototype screens), the user may connect to other applications and devices, and also mark trails and individual items. The menu may be used to view tagged/marked trails. These features may also be accessed by a context menu as shown in Figure 4-14.

One of the experts suggested that we include time and date in the main screen of the display so the Personal Memex could be used as a watch in the same way that cell phones are used to check current time and date.

To help users with macular degeneration and those with MCI, one suggestion was to have mouse-over magnification of the display. This feature would magnify parts of the display as the user moves her mouse over it.

Some experts felt that we should use more prototypical or general terms on the screens. For example, the term "default" may be replaced by "standard". One solution could be to elicit this information from the user. As with text, image and sound, we can allow the user to set preferences for the terms that the user would like to see (that represent different functions) on the user interface. For example, some people may prefer to label the "Google" button on the search toolbar as "Search World" (although in the background the system may still be searching using Google).

Other suggestions by the experts included the use of color to indicate piles of information already searched or visited, a help menu item incase the user needs assistance to use the Personal Memex, including the "Default" feature in the Find screens, and providing a facility to scroll forward so that the user may get information on future events.

## 4.2 Personal Memex Design Guidelines

Through our literature research and Cognitive Walkthroughs, we have collected the following recommendations, which we pose as guidelines for Memex functionalities and designs.

### 4.2.1 General Results

- Maintain consistent, meaningful graphics (Tognazzini)

- Support opportunistic navigation
  - user control and freedom (Neilsen; Tognazzini)
  - 'Bookmarking' paths, previous searches
- Leverage user recognition rather than recall (Neilsen) –
  - provide context for users while searching and browsing memories
- Provide visibility for mode & system status (Neilsen; Tognazzini)
  - To support situational awareness, provide visibility of current time and date whenever browsing, searching or scheduling in time
- Provide Internationalized Tool Tips for icons and buttons
- Provide Help w/o loosing state
- Provide Tutorials explicitly demonstrating functionality
- Since a disability can change over time or have a wide range of effects, support customization or adaptation of the interface.

### 4.2.2 Population-specific Results: Mild Cognitive Impairment
- Simplify task steps
- Provide views that can represent time and objects graphically
- Provide time, space, and object Reminders (for schedules, directions, medication)

### 4.2.3 Population-specific Results: Visual Impairment
- Provide enlargeable fonts & interface items (e.g. buttons)
- Provide contrast between font and background
- Provide options for audio output: voice, speed, pitch

## 4.3 Conclusions

The goal of this project was to consider the use of a Personal Memex by several populations, the high-functioning population, the population with mild cognitive impairment and individuals suffering from macular degeneration.  By creating a customizable, adaptable set of preferences, which could be adjusted by the user or a caregiver, we were able to support each unique population, while at the same time, providing the same universal design for the rest of the Personal Memex.

We interviewed experts, developed scenarios, and drew a low-fidelity prototype.  We then went back to our experts with our prototype and conducted walkthroughs.  Our experts enthusiastically and positively received our prototype.  To this author, our design was validated when, midway through the walkthrough, one expert said, "This is cool!  I want one of these!"  (Roberto, 2005)

# Appendix A    Project Proposal

**Finding and Using Digital Memories:
Human Information Processing
with the Personal Memex**

**Semester Project
ISE 5604: Human Information Processing
Fall 2005**

**Project Client**

Dr Manuel Perez-Quinoñes

_______________________________________

<span style="font-size:smaller">Signature</span>

**Project Team**

| Name | E-mail Address | Signature |
|------|----------------|-----------|
| Nicholas Polys | npolys@vt.edu | |
| Uma Murthy | umurthy@vt.edu | |
| Ingrid Burbey | iburbey@vt.edu | |
| Gyuhyun Kwon | ghkwon@vt.edu | |
| Prince Vincent | princev@vt.edu | |

# Problem statement

Vannevar Bush envisioned a future where a "Memex" device would categorize and organize information for the professional. For this project, we will explore the Human Information Processing aspects of a Personal Memex, which is a Memex to organize personal (instead of professional) information. This exploration will result in an annotated bibliography, a low-fidelity prototype and a set of information design guidelines for such a system.

# Background Information

The idea of a "Memex" was publicly proposed in 1945. Advances in digital storage, information retrieval, internetworking and wearable computing have made the idea into a possibility. Microsoft researchers have established the MyLifeBits project, which records users' daily activities. Similar projects by other researchers include the Haystack project at MIT and Keeping Found Things Found at University of Washington.

Our project assumes that the each user has a SenseCam, a device developed by Microsoft, which captures pictures every few seconds and records other data, such as audio, temperature and GPS readings. We also assume that the user's personal electronic information, such as emails, calendar, personal files, and phone logs are also available. All of this information is magically organized and indexed without any human effort and is easily retrievable.

# Project Goals

This work fits into a larger research agenda on Personal Space of Information (PSI), pervasive computing and the use of small form-factor wearables such as the SenseCam. Our work will apply Usability Engineering principles to the design of a personal, ubiquitous digital memory system. Our goal is not to design some sort of universal memory aid, but to consider the issues involved and discover the resulting design guidelines.

We target 3 populations: a 'high-functioning group (those with no apparent disabilities), those diagnosed with Mild Cognitive Impairment (MCI), and those diagnosed with Macular Degeneration. Targeting an MCI population provides us with an opportunity to research ways the Memex interface can be adapted with features that can aid the completion of short term goals. The Macular Degeneration population gives us reason to consider accessibility and alternate rendering features that can increase operator-system bandwidth under poor eyesight conditions.

Through the models of Human Information Processing and Engineering Psychology, we will derive a set of features, claims, and design guidelines about this system. We will examine various cognitive styles and organizing principles; for example, patterns of associations that users employ in their mental models.

# Method

We will use scenario-based design, cognitive walkthrough, and literature research to complete our goals and deliverables (listed below). We will derive initial requirements through scenarios and develop a prototype. We will conduct a cognitive walkthrough on this initial prototype with population experts, Dr. Karen Roberto and Dr. Richard Snyder. Using feedback and results from this process, we will revise our scenarios, and hence our prototype and design guidelines. We will use literature research to guide us through various stages of the project.

Our scenarios may focus on two novel classes of people – musicians and amateur athletes. To create a needs assessment, we will apply these scenarios to each of our populations to develop a low-fidelity prototype. An example scenario could include time management. Recommendations for organizing time could be based on associations such as best time of the day and environment to perform creative tasks (for musicians) or to practice (for both musicians and athletes). A direct benefit of being able to recall a memory includes the opportunity to replay key performances (musical or athletic) or to remember instructors' or trainers' instructions.

In addition to scenarios, we will consider, for each population in each class, pertinent human information processing principles such as:
- Signal detection theory, information theory, attention, mental models, population stereotypes,
- M-level activities (maintenance and organization),
- Decisions on keeping and discarding information (finding & re-finding), and
- Interfaces for navigating and displaying memory information.

Our design guidelines will be applicable to several device platforms. However, our prototype will be based on one or two platforms driven by our design guidelines.

# Analyses

Guiding us in this process are the models of Human Information Processing, which we will use as initial guidance for design. Finally, we will use an analytic inspection method, cognitive walkthrough, to assess the prototype and provide claims evidence for future designs. Along the way, we will formalize and record any trade-offs encountered as well as a set of features and claims; each claim will be justified by references to the annotated bibliography.

# Deliverables
- HIP-centered design requirements list and design guidelines for the interface and functionality that will make the device effective for both those with disabilities and the general population,
- Low-fidelity prototype (screen mockups and story boards),

- Annotated bibliography including HIP, usability, and design literature related to the project,
- Project report.

# Time Line & Milestones

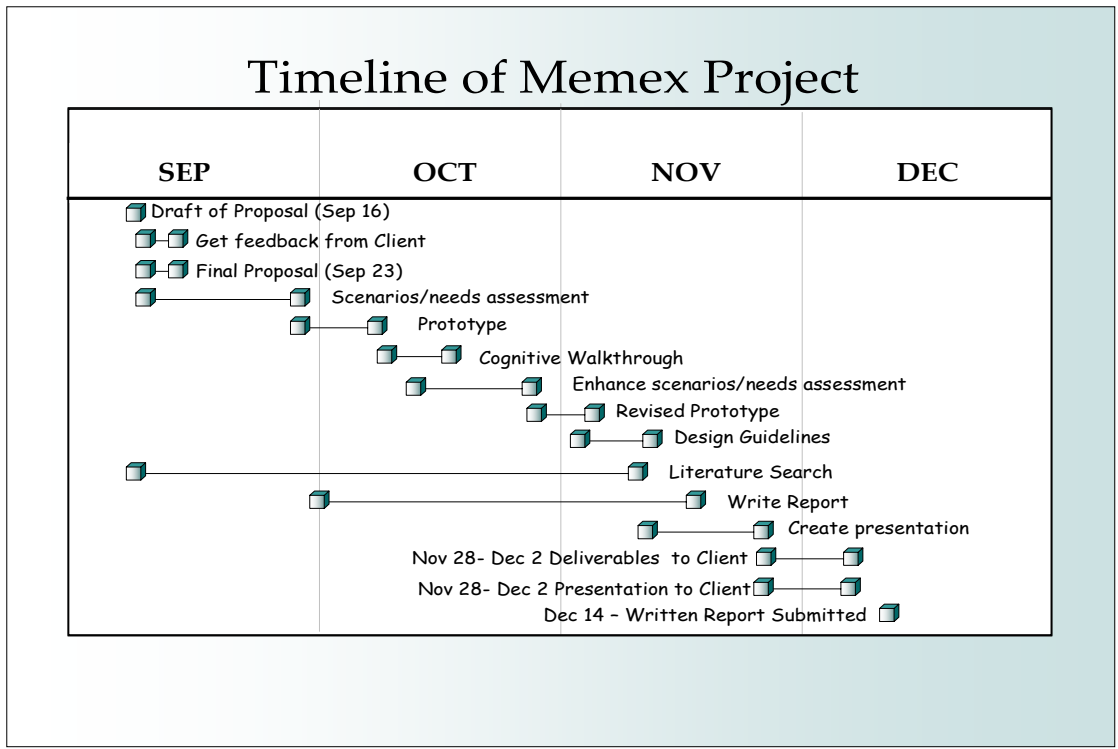

# Team Members: Qualifications & Roles

| Member | Qualifications | Interests and Experience | Role[*] |
|---|---|---|---|
| Nicholas Polys | B.A. Cognitive Science, 7 years Industry | Information-rich virtual environments, internet publication systems, and information design. | Design guidelines |
| Uma Murthy | M.S. (Computer | Superimposed | Prototype |

---

[*] This is not the distribution of tasks. It indicates the person responsible for this part of the project.

| | Applications) | information, digital libraries, and software agents. | |
|---|---|---|---|
| Ingrid Burbey | M.S. ECE | Pervasive computing. | Report and presentation |
| Gyuhyun Kwon | M.S. Industrial Engineering | Cognitive systems engineering, HCI, and game design. | Scenarios and cognitive walkthrough |
| Prince Vincent | B.S. Mechanical Engineering | Industrial systems and manufacturing. | Literature research |

# Appendix B    Annotated Bibliography

## B.1    Long term Memory

**Fenker, Daniela B.; Schott, Björn H.; Richardson-Klavehn, Alan; "Recapitulating emotional context: Activity of amygdala, hippocampus and fusiform cortex during recollection and familiarity." European Journal of Neuroscience, Vol 21(7), Apr 2005, pp. 1993.**

The amygdala is thought to enhance long-term memory for emotionally arousing events by modulating memory formation and storage in the hippocampus and in neocortical areas. The objective of this study was to see if there is any relation between the amygdale and the hippocampus during the retrieval of emotional memories. In this functional magnetic resonance imaging(fMRI) experiment, 20 healthy subjects studied neutral words in the context of a fearful or a neutral human face. Subsequently, they made 'remember' (conscious recollection of the study context), 'know' (familiarity in the absence of conscious recollection) and 'new' judgements on the studied and newly presented neutral words, in the absence of face stimuli. From the fMRI, it was observed that bilateral amygdala, hippocampus and fusiform face area (FFA) were more strongly activated during recollection than during familiarity. Higher activity was found for fearful than for neutral study context during recollection but not during familiarity. The study suggests that the amygdalae and hippocampi contribute to the retrieval of emotion-laden context memories.

**Anderson, Michael C.; Bjork, Elizabeth L.; Bjork, Robert A.; "Retrieval-induced forgetting: Evidence for a recall-specific mechanism." Psychonomic Bulletin & Review, Vol 7(3), Sep 2000, pp. 522.**

Previous works have shown that recalling information from long-term memory can impair the long-term retention of related representations--a phenomenon known as retrieval-induced forgetting (M. C. Anderson et al, 1994). This study tries to ascertain the validity of the above premise based on a series of tests. These findings points to the fact that retrieval-induced forgetting is not caused by increased competition arising from the strengthening of practiced items, but by inhibitory processes specific to the situation of recall.

**Tulving, Endel; Kroll, Neal; "Novelty assessment in the brain and long-term memory encoding." Psychonomic Bulletin & Review, Vol 2(3), Sep 1995, pp. 387.**

The purpose of this study was to test the hypothesis that the efficacy of encoding on-line information into long-term memory depends on the novelty of the information as determined by novelty-assessment networks. A test of this "novelty/encoding" hypothesis was conducted with 4 university-educated adults. The subjects studied a list of words. Half of the words were "familiar" the other half were novel. Results support the novelty/encoding hypothesis: Accuracy of explicit (episodic) recognition was higher for novel than for familiar words.

**Sutherland, Rachel; Pipe, Margaret-Ellen; Schick, Katherine; "Knowing in advance: The impact of prior event information on *memory* and event knowledge." Journal of Experimental Child Psychology, Vol. 84(3), Mar 2003, pp. 244.**

This study was aimed at determining whether prior information related to an upcoming event can alter the way information is being processed in the memory. Thirty-five children between the ages of 5 and 7 years participated in the novel event (Visiting the Pirate). The day before participating, children were: (1) provided with new information specific to the up-coming event; (2) engaged in a discussion generally related to the event topic based on existing knowledge; or (3) discussed an unrelated topic. It was found that advance information specific to the event led to better recall and, in particular, to better integration of the experience into a general event representation both soon after the event and at a follow-up interview 4 months later. A general discussion of the topic without the event specific information didn't enhance the memory processing. It can be observed form this study that providing information in advance can have significant effects on memory.

**Winkler, István; Cowan, Nelson; "From Sensory to Long-Term Memory: Evidence from Auditory Memory Reactivation Studies." Experimental Psychology, Vol. 52(1), 2005, pp. 3.**

Even though it is known that auditory sensory memory doesn't hold anything for more than a few seconds. But at the same time wee seem to remember people based on their voices. This study tries to bring more insight into this process of transferring of auditory information into long tem memory by focusing on the following questions.
(1) What types of acoustic information can be retained for a longer term?
(2) What circumstances allow or help the formation of durable memory records for acoustic details?
(3) How such memory records can be accessed?
The paper discusses the results of experiments that used a model of auditory recognition, the auditory memory reactivation paradigm. Results obtained with this paradigm suggest that the brain stores features of individual sounds embedded within representations of acoustic regularities that have been detected for the sound patterns and sequences in which the sounds appeared. Thus, sounds closely linked with their auditory context are more likely to be remembered. The representations of acoustic regularities are automatically activated by matching sounds, enabling object recognition.

## B.2 Episodic & Autobiographical Memory

**Mayes, Andrew R.; Roberts, Neil; "Theories of episodic memory. Episodic memory: New directions in research." Baddeley, Alan (Ed); Aggleton, John P. (Ed); Conway, Martin A. (Ed); pp. 86.**

The main focus in this paper is on three theoretical characteristics of episodic memory: (1) the encoding, storage, and retrieval processes that mediate this form of memory; (2) the brain regions that perform each of these processes and how these regions interact with each other so that the processes work in an integrated manner; and (3) how these episodic memory processes are similar to and/or differ from the processes that mediate other forms of memory, and to what extent the same brain regions mediate memory for episodes and other kinds of information. The process and information involved in making episodic representations at encoding are outlined and the underlying brain regions are considered. Major views about the psychological and physiological processes underlying consolidation and long-term storage of episodic representations are outlined and differing views concerning where these processes are located in the brain are discussed. Views about how retrieval is achieved and the feeling that retrieved episodes are being remembered, as well as which brain structures mediate these processes, are discussed. Similarities and differences between episodic memory and both semantic memory and priming are also discussed.

**Anderson, Stephen J.; *Conway*, Martin A. "Representations of autobiographical memories."Conway, Martin A. (Ed); pp.217.Cambridge, MA, US: The MIT Press, 1997**

The authors focus on how adequately the different approaches treat the representation of specific experiences, i.e., autobiographical memories (AMs). Three broad "schools" of knowledge representation are discussed: semantic networks, scripts, and connectionist models. The ability of the 3 approaches to account for a critical list of facts is appraised. The final section briefly considers issues relating to the roles of the self and central control processes in mediating remembering.

**Hayes, Scott M.; Ryan, Lee; Schnyer, David M.; Nadel, Lynn; "An fMRI Study of Episodic Memory : Retrieval of Object, Spatial, and Temporal Information." Behavioral Neuroscience. Vol. 118(5) October 2004. pp. 885.**

The goal of the present study was to address the regional patterns of activation in frontal and medial temporal lobe structures during recollection of specific elements of episodic memory. The study mailf focused on the three defining components of episodic memory, namely content (objects or items), spatial context, and temporal information.. Participants were tested for memory of object, spatial, and temporal-order information while undergoing functional magnetic resonance imaging. Preferential activation was observed in the right parahippocampal gyrus during the retrieval of spatial-location information. Retrieval of contextual information (spatial location and temporal order) was associated with activation in the right dorsolateral prefrontal cortex. In bilateral posterior parietal regions, greater activation was associated with processing of visual scenes regardless of the memory judgment. These findings support current theories positing roles for frontal and medial temporal regions during episodic retrieval and suggest a specific role for the hippocampal complex in the retrieval of spatial-location information.

**Souchay, Céline; Isingrini, Michel; Espagnet, Laurence; "Aging, Episodic Memory Feeling-of-Knowing, and Frontal Functioning," Neuropsychology. Vol. 14(2) April 2000, pp. 299.**

Groups of normal old and young adults made episodic memory feeling-of-knowing (FOK) judgments and took 2 types of episodic memory tests (cued recall and recognition). Neuropsychological tests of executive and memory functions thought to respectively involve the frontal and medial temporal structures were also administered. Age differences were observed on the episodic memory measures and on all neuropsychological tests. The finding of the study supports the idea that the age-related decline in episodic memory FOK accuracy is mainly the result of executive or frontal limitations associated with aging.

**Ellis, Henry C.; Thomas, Roger L.; McFarland, Alan D.; Lane, J. Walter; "Emotional Mood States and Retrieval in Episodic Memory."Journal of Experimental Psychology: Learning, Memory, and Cognition. Vol. 11(2) April 1985, pp. 363.**

The objective of the study was to determine whether the induction of a depressed mood can affect output or retrieval from episodic memory. The experimental sequence was as follows: All subjects studied a list of either elaborated or base sentences, rating them for complexity, in an incidental retention paradigm. This was followed by the induction of a depressed or neutral (control) mood. Then the subjects were given an unanticipated cued recall test of the target adjectives. In all tests, subjects showed a reduction in recall owing to the depressed mood, which provided evidence for retrieval effects of the mood state. The results are briefly discussed within the framework of a resource allocation theory.

**Hultsch, David F.; MacDonald, Stuart W. S.; Hunter, Michael A.; Levy-Bencheton, Judi; Strauss, Esther; "Intraindividual Variability in Cognitive Performance in Older Adults : Comparison of Adults With Mild Dementia, Adults With Arthritis, and Healthy Adults." Neuropsychology. Vol. 14(4) October 2000, pp. 588.**

The study aimed at evaluating the intraindividual variability in latency and accuracy of cognitive performance across both trials and occasions was examined in 3 groups of older adults: healthy adults, adults with arthritis, and adults diagnosed with mild dementia. Participants completed 2 reaction-time and 2 episodic-memory tasks on 4 occasions. Results indicated that intraindividual variability in latency was greater in individuals diagnosed with mild dementia than in adults who were neurologically intact, regardless of their health status. .

**Allen,P.A., Kaut,K.P., Lord,R.G., Hall,R.J., Bowie,T., & Grabbe,J "An Emotional Mediation Theory of Differential Age Effects in Episodic and Semantic Memories." Experimental Aging Research, Vol 31(4), Oct-Dec 2005, pp. 355.**

It is a known fact that, although there is a large decrement in central episodic memory processes as adult's age, there is no appreciable decrement in central semantic memory processes. This paper discusses the theory behind this phenomenon within the context of the cognitive neuroscience research regarding limbic system connectivity. The central hypothesis is that elements of limbic system circuitry, including portions of the medial temporal lobes and frontal cortex, are associated with both working and long-term episodic memory performance, and by extension, with the capacity to engage in emotion-guided, self-regulatory processes that depend heavily on episodic memory. In contrast, the semantic memory system may have less shared interface with episodic and affective networks (i.e., the limbic-related system), and therefore remain independent of neurocognitive changes impacting emotional states and episodic-type memory processes.

## B.3  Associative Memory

**Achim, Amélie M.; Lepage, Martin**. "**Neural Correlates of Memory for Items and for Associations: An Event-related Functional Magnetic Resonance Imaging Study**." **Journal of Cognitive Neuroscience 17, no. 4 (2005): 652-667.**

The objective of the study was to get some insight into the function of the various parts of the brain related to associative memory. The researchers used event-related functional magnetic resonance imaging to assess neural correlates of item and associative encoding and retrieval. During encoding, subjects memorized items and pairs. During retrieval, subjects made item recognition judgments and associative recognition judgments.
Relative to baseline, item and associative trials activated bilateral medial temporal and prefrontal regions during both encoding and retrieval. During encoding, it was observed that there was greater prefrontal, hippocampal, and parietal activation for associations, but no significant activation was observed for items at the selected threshold. For recognition, greater activation was observed for associative trials in the left dorsolateral prefrontal cortex and superior parietal lobules bilaterally, whereas item recognition trials showed greater activation of bilateral frontal regions, bilateral anterior medial temporal areas, and the right temporo-parietal junction.

**Touryan, Sharon Roseanna** "**The effect of emotional stimuli on memory for contextual information**" **Disseratation Abstracts Online,DAI, 66, no. 02B (2004): p. 1193**

The purpose of this study was to obtain more information about the following questions.
 (1) How the emotionality of item information influences the association of item and peripheral information in memory?
(2) How an external arousing event (an event separate from attended-to information) influences the association of item and central feature information in memory?
(3) How the emotionality of item information influences memory for temporal context?
 Findings indicated that item information is improved by emotional arousal regardless of the locus of the arousing event. It was observed that associative memory for peripheral information was reduced by emotional arousal, regardless of the locus of the arousing event or the spatial location of the peripheral information relative to the item information. For temporal context, according to the paper, memory is influenced by changes in emotional arousal rather than absolute emotionality.

**Collie, Alexander; Myers, Catherine; Schnirman, Geoffrey, et al. "Selectively Impaired Associative Learning in Older People with Cognitive Decline" Journal of Cognitive Neuroscience 14, no. 3 (2002): 484.**

Older people with declining cognitive function typically display deficits in associative learning (AL). The hippocampal formation (HF) is involved in the encoding and retrieval of such associations. The study involved a set of experiments to determine the cause and nature of difficulties such population has with respect to associative learning. The results of the study pointed to the fact that older people with cognitive decline displayed impaired performance on tasks designed to be highly dependent upon intact hippocamapal formation function, including a task in which novel patterns and spatial locations were to be associated. The result from this paper suggests that the associative learning impairments observed in older people with cognitive decline may be due to HF dysfunction.

## B.4   Short-Term Memory

**Haarmann, Henk; Usher, Marius; Maintenance of semantic information in capacity-limited item short-*term memory* Psychonomic Bulletin & Review, Vol 8(3), Sep 2001. pp. 568-578.**

The authors report a semantic effect in immediate free recall, which is localized at recency and is preserved under articulatory suppression but is highly reduced when recall is delayed after an intervening distracter task. They also suggest that in addition to the phonological component in verbal short term memory, there is an activation/item-limited component with semantically sensitive representations.

**Freedman, Monica L.; Martin, Randi C.; "Dissociable components of short-term memory and their relation to long-term learning." Cognitive Neuropsychology, Vol 18(3), 2001. pp. 193.**

The objective of this study was to ascertain the effects of semantic and phonological components of STM in the process of long term learning.  Five aphasic patients were tested who to see how deficits in the short-term retention of either phonological or semantic information affected the process of long term leaning. From the results of the study, it can be inferred that semantic and phonological components of STM are essential for the long-term learning.

**Akyürek, Elkan G.; Hommel, Bernhard; "Short-term memory and the attentional blink: Capacity versus content. " *Memory* & Cognition, Vol 33(4), Jun 2005, pp. 654.**

When people monitor the rapid serial visual presentation (RSVP) of stimuli for two targets, they often miss the second one if it falls into a time window of about half a second after the onset of the first one. This phenomenon is known as the attentional blink (AB). This study found that overall performance in an RSVP task was impaired by a concurrent short-term memory (STM) task. This effect increased when STM load was higher and when its content was more task relevant. Loading visually defined stimuli and adding articulatory suppression further impaired performance on the RSVP task.

**Todd, J. Jay; Marois, René; "Posterior parietal cortex activity predicts individual differences in visual short-term memory capacity." Cognitive, Affective & Behavioral Neuroscience, Vol 5(2), Jun 2005, pp. 144.**

Humans show a severe capacity limit in the number of objects they can store in visual short-term memory (VSTM). People vary widely in their VSTM capacity. In this study, the researchers examine the neural basis of these individual differences. An individual-differences analysis revealed a significant correlation between posterior parietal cortex (PPC) activity and individuals' VSTM storage capacity. They also suggested that other parts of the brain, particularly visual occipital cortex, may contribute to individual differences in VSTM capacity. The findings from this study support an important role for the PPC in visual short-term memory (VSTM).

**Purser, Harry R. M.; Jarrold, Christopher; "Impaired verbal short-term memory in Down syndrome reflects a capacity limitation rather than atypically rapid forgetting." Journal of Experimental Child Psychology, Vol 91(1), May 2005. pp. 1**

Individuals with Down syndrome suffer from relatively poor verbal short-term memory. There is an opinion that this is not caused by problems of audition, speech, or articulatory rehearsal within the phonological loop component of Baddeley and Hitch's working memory model. This research attempts to analyze some of the factors that can be the reason for the poor verbal short term memory in Down syndrome. Two experiments were conducted to investigate whether abnormally rapid decay is the basis this deficit. The results indicated that this may be due to a limited-capacity verbal short-term memory system.

## B.5        Working Memory

**Kiefer, Markus; Ahlegian, Michelle; Spitzer, Manfred; "Working Memory Capacity, Indirect Semantic Priming, and Stroop Interference: Pattern of Interindividual Prefrontal Performance Differences in Healthy Volunteers "Neuropsychology, Vol 19(3), May 2005. pp. 332.**

In this study the authors compared verbal and visuospatial working memory capacity with performance on the Stroop interference task as a measure of executive control. They compared direct and indirect semantic priming tasks as a measure of semantic access. It was found that people with low visuospatial working memory capacity exhibited increased Stroop interference. People with low verbal memory capacity showed increased priming and Stroop interference. These findings from this study suggest that working memory, executive control, and semantic retrieval are functionally related to some extent.

**Ranganath, Charan; Cohen, Michael X.; Brozinsky, Craig J.; "Working *Memory* Maintenance Contributes to Long-term *Memory* Formation: Neural and Behavioral Evidence." Journal of Cognitive Neuroscience, Vol 17(7), Jul 2005, pp. 994.**

In this study, the researchers tested the hypothesis that working memory (WM) maintenance operates in two stages, and that processing during the initial stage of working memory maintenance promotes successful long term memory (LTM) formation. Results from a functional magnetic resonance imaging study showed that activity in the dorsolateral prefrontal cortex and hippocampus during the initial stage of working memory maintenance was predictive of subsequent long term memory (LTM) performance. In a behavioral experiment, it was demonstrated that interfering with processing during the initial stage of working memory (WM) maintenance impaired long term memory (LTM) formation. These results suggests that processing during the initial stage of working memory (WM) maintenance directly contributes to successful long term memory (LTM) formation.

**Oberauer, Klaus; "Access to information in working memory: Exploring the focus of attention." Journal of Experimental Psychology: Learning, Memory, and Cognition, Vol 28(3), May 2002, pp. 411.**

In this paper the researcher proposes a model of working memory that can be considered as being made of 3 functionally distinct regions with the following functions.

1. The activated part of long-term memory that can memorize information over brief periods for later recall.
2. The region of direct access that can hold a limited number of chunks available to be used in ongoing cognitive processes.
3. The focus of attention that can hold time the one chunk that is actually selected as the object of the next cognitive operation.

The author derived these above results based on 2 sets of experiments he conducted. He also presents some interesting factors affecting the processing of information in working memory.

**Samman, Shatha N.; "Multimodal computing: Maximizing working memory processing." Dissertation Abstracts International: Section B: The Sciences and Engineering, Vol 65(9-B), 2005. pp. 4864.**

According to multiple resource theory, enhancements in human information management capacity can be realized using multimodal interaction.. This study proposes an expansion of the current bimodal (verbal, visual/spatial) model of working memory(WM) to a multimodal WM system, which includes different modalities like verbal, visual, spatial, kinesthetic, tactile, and tonal. The researcher carried out a set of experiments to study the various facets of multi modal computing. These tests compared the memory performance based on various combinations of multiple modalities against single modality. Form these experiments it was inferred that in certain combinations multimodal WM capacity averaged more than three times the 'magic number' seven proposed by Miller's (1956) unidimensional memory span. It was observed that some times multimodal capacity nearly reached the summation of each single modality capacity.

It was demonstrated that there is minimum interference between the subsystems in terms of storage and rehearsal mechanisms. Some theoretical and applied implications for the use of multimodal interaction are also discussed in this paper.

**Wagar, Brandon M.; Dixon, Mike J.; "Experience influences object representation in working memory." Brain and CognitionPast , Vol 57(3), Apr 2005. pp. 248.**

This study involved two experiments to test whether a diagnosticity effect exists in working memory; and whether it is present when visual information is encoded into working memory, or if it is the result of maintenance within working memory. Results showed a diagnosticity effect was present at encoding. The results from this study, directs to the fact that the meaning we glean from our past experience has a profound influence on the nature of object representation in working memory.

## B.6   Cognitive Impairment

**Kazui, Hiroaki; Matsuda, Akemi; Hirono, Nobutsugu. "Everyday Memory Impairment of Patients with Mild Cognitive Impairment." Dementia and Geriatric Cognitive Disorders, Vol 19(5-6), May 2005. pp. 331-337.**

The authors evaluated the memory impairment in 24 patients with mild cognitive impairment (MCI) using the Rivermead Behavioral Memory Test (RBMT) to determine how MCI affects their everyday tasks. The scores were then compared with those of 48 similar Alzheimer disease (AD) patients and 48 normal controls (NC). It was inferred from this study that the level of impairment in the MCI population was less severe than the AD patients. The MCI population showed impairment with respect to tasks requiring delayed recall. But they could perform these tasks immediately after memorizing except for recalling and retracing a simple new route.

**Tales, Andrea; Snowden, Robert J.; Haworth, Judy. "Abnormal spatial and non-spatial cueing effects in mild cognitive impairment and Alzheimer's disease." Neurocase, Vol 11(1), Feb 2005. pp. 85-92.**

The paper examines the visual-attention related processing capabilities of three different groups, one group with mild cognitive impairment (MCI), the other group with Alzheimer's disease (AD) and the normal population. The study suggests that the MCI group and the AD group exhibited a significant detriment in both the ability to disengage attention from an incorrectly cued location and the ability to use a visual cue to produce an alerting effect.

**Tabbers H.K.; Martens R.L.; van Merriënboer J.J.G. "Multimedia instructions and cognitive load theory: Effects of modality and cueing."  British Journal of educational Psychology, Volume 74, Number 1, March 2004, pp. 71-81.**

Both cognitive load theory (Sweller, Van Merriënboer, & Paas, 1998) and Mayer's theory of multimedia learning (Mayer, 2001), suggests that replacing visual text with spoken text and adding visual cues would increases the effectiveness of multimedia instructions in terms of better understanding and less mental effort spent. In this paper, the authors examine the validity of the above statement. 111 participants studied a web-based multimedia lesson on instructional design for about an hour and were asked to complete a retention and transfer test. It was concluded that adding visual cues to the pictures results in higher retention scores, while replacing visual text with spoken text results in lower retention and transfer scores.

**Leritz, Elizabeth C. "Associative priming and explicit memory in aging and mild cognitive impairment." Dissertation Abstracts International: Section B: The Sciences and Engineering, Vol 65(8-B), 2005, pp. 4293.**

The main objective this study was to study the effects mild cognitive impairment on explicit and implicit priming. The participants completed an associative priming experiment which assessed implicit associative memory for novel and semantic word pairs. Following the priming experiment, all participants completed an incidental cued-recall test for the novel and semantic word pairs previously seen. Results revealed that MCI and control groups did not differ much on novel and semantic priming, or on incidental cued-recall. From the study it can be inferred that MCI impairs explicit but not implicit associative memory.

**Xiaoming, Zhang; Geng, Xu; Qihao, Guo. "Cognitive Hold and Deficit of Patients with Mild Cognitive Impairment." Chinese Mental Health Journal, Vol 18(10), Oct 2004. pp. 685.**

This study analyzed the cognitive hold and other cognitive deficits in patients with mild cognitive impairment (MCI). A battery of memory tests were conducted on three population groups, one group with mild cognitive impairment (MCI), the other group with Alzheimer's disease (AD) and the normal population. It was concluded that there was a significant difference in the memory functions between the normal population and the mild cognitive impairment group. But the two groups did not differ much in attention, initiation/perseverance, construction and conceptualization subscales. The MCI group scored significantly higher than the AD group in terms of total intelligence, memory, verbal fluency and executive function.

**Moulin, Christopher J. A.; James, Niamh; Freeman, Jayne E. "Deficient acquisition and consolidation: Intertrial free recall performance in Alzheimer's disease and mild cognitive impairment." Journal of Clinical and Experimental Neuropsychology, Vol 26(1), Feb 2004. pp. 1-10.**

It ha been known that dementia of the Alzheimer type (DAT) and mild cognitive impairment (MCI) are associated with deficits in episodic memory, especially with respect to acquisition of new information. Some researchers have argued that there is also a deficiency in the consolidation of new information in this population. This study tries to see if there is any real correlation between the acquisition and consolidation of information in these groups. The results of the study did not find any correlation between the acquisition and consolidation within any of the groups, which suggests that there may not be a direct correlation between the acquisition and consolidation of data.

**R. C. Petersen. "Mild cognitive impairment as a diagnostic entity." Journal of Internal Medicine Volume 256 Issue 3 Page 183 -194 September 2004.**

Cognitive impairment has always been thought as a n intermediate stage between normal aging and dementia. Mild cognitive impairment refers to an early stage of cognitive impairment. This paper tries to bring to light the different types or classifications of mild cognitive impairment and comes with guidelines as how to differentiate between the various stages of mild cognitive impairment. The paper deals with amnestic MCI, non-amnestic MCI and the further sub-divisions.

**B. Winblad, K. Palmer, M. Kivipelto, V. Jelic, L. Fratiglioni et al. "Mild cognitive impairment – beyond controversies, towards a consensus: report of the International Working Group on Mild Cognitive Impairment."**

The aim of the symposium was to integrate clinical and epidemiological perspectives on the topic of Mild Cognitive Impairment (MCI). Experts from multiple domains related to cognitive impairment discussed the current status and future directions of MCI, with regard to clinical presentation, cognitive and functional assessment, and the role of neuroimaging, biomarkers and genetics. The specific recommendations for the general MCI criteria include the following from the symposium were:
  (i)     The person is neither normal nor demented.
  (ii)    There is evidence of cognitive deterioration shown by either objectively measured decline over time and/or subjective report of decline by self and/or informant in conjunction with objective cognitive deficits.
  (iii)   Activities of daily living are preserved and complex instrumental functions are either intact or minimally impaired.

**Whitlatch, Carol ; Feinberg, Lynn ; Tucke, Shandra. "Accuracy and consistency of responses from persons with cognitive impairment." Dementia 4, no. 2 (2005): 171-183.**

The purpose of the study was to examine the ability of people with cognitive impairment to give accurate and consistent responses to questions about demographic characteristics and basic preferences. People with cognitive impairment were interviewed twice within one week in order to determine stability and accuracy of responses. Family caregivers were interviewed once within the same time period. It was observed that persons with mild cognitive impairment were accurate and reliable in their ability to respond to questions about demographics and basic preferences.

**Perri, R.; Carlesimo, G. A.; Serra, L., et al. "Characterization of Memory Profile in Subjects with Amnestic Mild Cognitive Impairment." Journal of Clinical and Experimental Neuropsychology 27, no. 8 (2005): 1033-1055**

This paper investigated the several aspects of memory; episodic long-term, short-term and implicit long-term associated with the amnestic Mild Cognitive Impairment (a-MCI) population. Results showed that the subjects possessed normal short-term memory abilities. But factors related to episodic long-term memory, showed poorer results in a-MCI subjects with respect to normal controls. Some episodic memory functions were relatively well preserved, while others appeared to have degenerated a lot.

**P. Maruff, A. Collie, D. Darby, J. Weaver-Cargin, C. Masters, Jon Currie. "Subtle Memory Decline over 12 Months in Mild Cognitive Impairment." Dementia and Geriatric Cognitive Disorders 18, no. 3-4 (2004): 342-348**

The objective of this study was to determine if mild cognitive impairment is a steady state or if it deteriorates to other kinds of cognitive impairments. A group of people with mild cognitive impairment (amnestic version) and another control group were studied over a period of 12 months to verify the changes in the cognitive abilities across this period. The groups were subjected to a continuous learning task (CLT), a word list learning task and a computerized paired associative learning task. The memory performance of the MCI group was found to be significantly worse than the control group, for all the memory tests. For the continuous learning task (CLT), the difference was found to be even more at the end of the 12 month monitoring period. This was due to the decline in memory accuracy in the MCI group. But no decline was observed in the case of routine memory tasks during the same period. Tus it can be inferred from this study that MCI is not a steady state condition.

**Crowell, Timothy A. ; Luis, Cheryl A. ; Vanderploeg, Rodney D. ; Schinka, John A. ; Mullan, Michael. "Memory Patterns and Executive Functioning in Mild Cognitive Impairment and Alzheimer's Disease." Aging, Neuropsychology, and Cognition 9, no. 4 (2002): 288-297**

The objective of the paper was to study the similarities in the pattern cognitive loss in Alzheimer's disease (AD) and Mild Cognitive Impairment (MCI). The factors under consideration were  memory acquisition, consolidation, and retrieval, as well as language, visuospatial ability, psychomotor speed, and executive functioning. Significant differences were observed in terms of consolidation of information among the 2 groups. No significant differences were observed in case of acquisition or retrieval.  The two groups also didn't differ in the pattern of memory processing.

**Cicerone, Keith. "Attention deficits and dual task demands after mild traumatic brain injury." Brain Injury 10, no. 2 (1996): 79-90**

Mild traumatic brain injuries (MTBI) often lead to cognitive dysfunctions. Attentional deficit seems to be the most important one among these. The purpose of this paper is to study the ability of this population to perform with background noise, and while simultaneously attending to a secondary task. The results showed that the dual task demand produced a significant slowing in processing speed for both the MTBI patients and control subjects. But the deterioration in processing speed appeared much greater for the patients with MTBI.

**Kirstie Hawkey, Kori M. Inkpen, Kenneth Rockwood, Michael McAllister, Jacob Slonim. "Designing for individuals with memory and cognitive disabilities: Requirements gathering**

**with Alzheimer's patients and caregivers**." **Proceedings of the 7th international ACM SIGACCESS conference on Computers and accessibility Assets '05**.

This paper delves in to the area of making use of technology to improve the life of people with cognitive disabilities. Repetitive questions by this population are a cause of stress for the caregivers. The authors describe the repetitive questions to be falling under 5 general categories-time, schedule, current event details, information and opinion/feedback. The paper identifies the main requirements any assistive device that can be of real help for this population should have. There are many constraints associated with this population group. So the assistive device should be one that can answer the repetitive kind of questions while accounting for the population constraints.

**Jessica Paradise, Elizabeth D. Mynatt, Cliff Williams, John Goldthwaite. "Designing a cognitive aid for the home: a case-study approach." Proceedings of the 6th international ACM SIGACCESS conference on Computers and accessibility. Pages: 140 – 146.**

This paper aimed at finding the issues related to building assistive technologies that can increase the functional independence of a cognitively impaired individual in the home environment. A case study approach was employed to find the needs for designing a pacing aid for an individual with a cognitive impairment. The contributions of this paper included insights gained with the two sets of design dimensions used: 'user-centered constraints' developed from capabilities and preferences of the users and 'system-centered capabilities' that could be explored in potential designs, a design concept which illustrates the application of these design dimensions into a potential pacing aid, and evaluations of paper prototypes guided by the design dimensions.

**Hendrik Schulze. "MEMOS: an interactive assistive system for prospective memory deficit compensation-architecture and functionality." Proceedings of the 6th international ACM SIGACCESS conference on Computers and accessibility (2004). Pages: 79 – 85.**

The paper focused on Mobile Extensible Memory Aid System (MEMOS), an electronic memory aid system which was developed to support patients with deficits in the prospective memory after a brain injury. The system consisted of a special palmtop computer, the Personal Memory Assistant (PMA), which reminded the patient of important tasks and supervises the patient's actions. The PMA communicated with a stationary care system via a bidirectional cellular radio connection (GPRS). MEMOS featured structured interactive reminding impulses, a flexible task scheduling, integration of a heterogeneous group of caregivers and integration in the patient's everyday life. The bidirectional communication allowed for reporting of critical situations back to a responsible care-giver, so MEMOS can be used even in a critical context. This paper described the requirements for such a memory aid system, the design and functionality of MEMOS as well as its application in the practical care of patients.

**Sri Kurniawan, Panayiotis Zaphiris. "Designing for individuals with memory and cognitive disabilities: Research-derived web design guidelines for older people." ACM SIGACCESS Conference on Assistive Technologies, Pages: 129 - 135 .Proceedings of the 7th international ACM SIGACCESS conference on Computers and accessibility.**

This paper presents a set of Web design guidelines for people with cognitive disabilities. The authors suggest that it is perhaps the first manuscript that proposes ageing-friendly guidelines that are for most part backed by published studies. The guidelines proposed in this study have been thoroughly examined through a series of expert and user verifications, which should give users of these guidelines confidence of their validity.

**Sharon Oviatt, Rachel Coulston, Rebecca Lunsford.** "**Multimodal interaction: When do we interact multimodally? Cognitive load and multimodal communication patterns**." **Proceedings of the 6th international conference on Multimodal interface Pages: 129 - 136 SESSION: Multimodal interaction**

This research investigated whether a multimodal interface would support users in managing cognitive load. Data from this study points to the fact that multimodal interface users spontaneously respond to dynamic changes in their own cognitive load by shifting to multimodal communication as load increases with task difficulty and communicative complexity. According to the4 authors this is accomplished by distributing communicative information across multiple modalities, which is compatible with a cognitive load theory of multimodal interaction. The data from this study supports the concept of multimodal communication in order to reduce cognitive load.

## B.7    Visual Impairment

**Holly S. Vitense, Julie A. Jacko, V. Kathlene, Emery. "Multimodal feedback: establishing a performance baseline for improved access by individuals with visual impairments." ACM SIGACCESS Conference on Assistive Technologies, Proceedings of the fifth international ACM conference on Assistive technologies, Pages: 49 – 56.**

This research investigated the effect in performance of fully sighted users while using unimodal, bimodal, and trimodal feedbacks. A complex direct manipulation task, consisting of a series search and selection drag-and-drop subtasks, was evaluated in this study. The different forms of feedback employed were auditory, haptic and visual. Each form of feedback was tested alone and in combination. User performance was assessed through measures of workload time. The results illustrated that multimodal feedback improves the performance of fully sighted users. These results point to the idea of using multimodal feedback mechanism while designing for the visually impaired.

**Lawton TA, Sebag J, Sadun AA, Castleman KR. "Image enhancement improves reading performance in age-related macular degeneration patients." Vision Research Volume 38, Issue 1 , January 1998, Pages 153-162.**

This study was aimed at determining if an image enhancement technique using filters specifically designed for patients with age-related macular degeneration (AMD) would be effective in improving the reading speed of scrolled text presented across the screen. The researchers found that this technique increased the average reading speed of the subjects by about 2.4 times. It was also observed that this enabled the reduction in magnification usually required for the AMD population. This study proposes the usefulness of individualized text filtering for image enhancement in digitally based viewing devices for the low vision group.

**Chandra M. Harrison. "Low-vision reading aids: reading as a pleasurable experience." Personal and Ubiquitous Computing Volume 8, Issue 3, Pages: 213 – 220.**

The paper discusses the relative advantages of myReader,a low-vision reading aid developed in New Zealand. They compare this with the most commonly used vision-aid for the visually impaired, the closed circuit television video magnifier (CCTV). Comfort ratings and preference results suggested that myReader provided a more pleasurable reading experience than traditional CCTVs. The study also highlights the aspects of myReader that should be altered to improve the experience by eliminating actions that cause errors and resulting negative emotions. Thus this study can be a helpful guideline as to what are the main features that should be included in designing vision-aids for the macular degeneration population.

**Steve Murphy. "Accessibility of graphics in technical documentation for the cognitive and visually impaired." Proceedings of the 23rd annual international conference on Design of communication: documenting & designing for pervasive information, SESSION: Graphical and visual information, Pages: 12 – 17.**

Even though the internet is a vast store of information, a lot of this information is not easily accessible to the disabled population. The ability to search and retrieve are major concerns for the cognitively and visually impaired population while accessing the internet. With lot of graphics being included in the internet, the Steve Murphy tries to evaluate the usability of internet from the perspective of the cognitively and visually impaired. The paper suggests two method of creating and exporting graphics that can improve the experiences of users with visual or cognitive impairments when viewing technical documentation:

- Clear, concise, and well-structured diagrams enable better comprehension for the cognitively impaired suffering from dyslexia and Attention-Deficit/Hyperactive Disorder (ADHD).
- The Scalable Vector Graphics (SVG) solution addresses many challenges for visually impaired people.

**Iain Murray, Helen Armstrong. "A computing education vision for the sight impaired." Proceedings of the sixth conference on Australian computing education - Volume 30, Pages: 201 – 206.**

This paper describes a research project undertaken by Curtin University in conjunction with Cisco Systems and the Association for the Blind to identify tools and techniques appropriate for vision-impaired students studying computing at tertiary level. It investigated learning characteristics of sight impaired students and considerations they may need. The paper discusses some of the teaching aids used to assist learning complex concepts usually delivered by visual means.

**Mary Zajicek. "Interface design for older adults." Proceedings of the 2001 EC/NSF workshop on Universal accessibility of ubiquitous computing, SESSION: Vision impairment and related assistive technologies, Pages: 60 - 65**

This paper evaluates the factors that seem to inhibit Web use by older adults, and explores aspects of human-computer interface design, which accommodate older users with age-associated disabilities. The paper concentrates on memory impairment, and cognitive and visual impairment. These conditions do not remain stable in older individuals. They vary from day to day and over longer time periods within an individual. The paper discusses a web browser specifically designed for the visually impaired, BrookesTalk. This was subsequently customized for older adults with memory loss. The researcher proposes the need for Design for Dynamic Diversity (DDD), an interface design approach, which accommodates design issues which come about as a result of changing user requirements related to older users' changing abilities.

**David A. Ross, Bruce B. Blasch. "Development of a Wearable Computer Orientation System." Personal and Ubiquitous Computing Volume 6, Issue 1 (February 2002) Pages: 49 – 63.**

The paper deals with the issue of aids for helping visually impaired people finding orientation/direction in their environment. Some of the requirements of such an orientation interface and the drawbacks of the existing systems are discussed. The authors specifically focused on the evaluation of three such wearable orientation interfaces: a virtual sonic beacon, speech output, and a shoulder-tapping system. The shoulder-tapping was found to be most useful. It can be inferred from this study that optimal performance and flexibility may best be obtained in a design that combines the best elements of both the speech and shoulder-tapping interfaces.

**Helen Petrie, Sarah Morley, Peter McNally, Anne-Marie O'Neill, Dennis Majoe. "Initial design and evaluation of an interface to hypermedia systems for blind users." Proceedings of the eighth ACM conference on Hypertext (1997), Pages: 48 – 56.**

Hypermedia provides great potential to overcome the difficulties that blind people have in accessing electronic information. This paper presents the initial designs for the hypermedia system which has a non-visual interface named DAHNI (Demonstrator of the ACCESS Hypermedia Non-visual Interface). DAHNI can be used with a variety of assistive input/output systems for blind users. Output from the system includes synthetic and digitized speech, non-speech sounds and refreshable Braille; input to the system can be via a small or large touch-tablet, joystick, and/or conventional keyboard. This paper presents an evaluation of DAHNI by seven blind and partially sighted students. Plans for further development based on these feedbacks are also discussed which can be used as guidelines while designing for the visually impaired.

**I. V. Ramakrishnan, Amanda Stent, Guizhen Yang. "Hearsay: enabling audio browsing on hypertext content." Proceedings of the 13th international conference on World Wide Web (2004), SESSION: Usability and accessibility, Pages: 80 – 89.**

This paper discusses HearSay, a system for browsing hypertext Web documents via audio. The HearSay system is based on automatically creating audio browsable content from hypertext Web documents. According to the authors it combines two key technologies: (1) automatic partitioning of Web documents through tightly coupled structural and semantic analysis, which transforms raw HTML documents into semantic structures so as to facilitate audio browsing; and (2) VoiceXML, an already standardized technology which is used to represent voice dialogs automatically created from the XML output of partitioning. This paper describes the software components of HearSay and presents an initial system evaluation.

**Vitense, Holly S. "Multimodal interface: Auditory, haptic, and visual feedback." Dissertation Abstracts International: Section B: The Sciences and Engineering, Vol 63(4-B), Oct 2002, pp. 2013.**

The study examined the effect of multimodal interface/feedback on the various performance measures like mental workload, accuracy etc. The researchers tested auditory, haptic, and visual feedback to assess their affect on user workload, accuracy, and performance time. Different combinations of these modalities were tested for the same tasks. The combinations used were (1) alone (i.e., unimodal), (2) combinations of two (i.e., bimodal), and (3) a combination of all three (i.e., trimodal). The results of the study suggest that certain types of multimodal feedback reduce mental workload and performance times, while increasing accuracy. In fact the study gives useful insight into which all combinations can be useful for the different population groups.

## B.8   Cognitive Prosthetics

**Cole, Elliot. "Cognitive prosthetics: An overview of function." NeuroRehabilitation, Vol 12(1), 1999, pp. 39-51**

This article can be termed as a general introduction to the field of cognitive prosthetics. It assesses the need for cognitive prosthetics in terms of gaps in rehabilitation methods and a response to managed care limitations on duration of services. The author tries to distinguish cognitive prosthesis from both electronic aids as well as from conventional computer software. Key researchers in this field as well as their accomplishments are discussed and also the differences in their approaches. The impact caused by advances in computers and communication devices on the field of cognitive prosthetics are also analyzed. This article concludes with a discussion of future research, noting that a key problem is the need to make cognitive prosthetics rehabilitation more user friendly to the therapist. This is considered necessary if cognitive prosthetics is to gain widespread use in rehabilitation facilities.

**Chaim-Meyer Scheff. "Experimental model for the study of changes in the organization of human sensory information processing through the design and testing of non-invasive prosthetic devices for sensory impaired people." ACM SIGCAPH Computers and the Physically Handicapped, Issue 36, Winter 1986, Pages: 3 – 10.**

The paper suggests a method of non-invasive prosthetic devices for sensory impaired people is proposed. These prosthetic devices are based on the transformation of inaccessible stimulus to impaired sensory modalities. Blind people would receive information about the visual world through their auditory senses and deaf people would receive information about the auditory world through their visual senses. The study accomplished the following objectives:

1. Build and tested a prosthetic device for the visually impaired.
2. Obtained objective measures of client adaptation to the devices.
3. Obtained subjective measures of the devices acceptance.
4. Monitored changes in the organization of the brain activity, due to the use of the proposed device.
5. Obtain new information for the modeling of the brain's organization.

**Elliot Cole , "Cognitive prosthetics: an overview to a method of treatment." Neurorehabilitation   Issue:  Volume 12, Number 1 / 1999   Pages:  39 - 51**

This paper discusses some of the factors that should be considered while designing tools for aiding cognitive prosthetics. Some of the suggestion from the paper while designing a system for cognitive prosthetics are:

(1) Patient priorities for activities
(2) Abilities in context of the environment where the target activity is per formed
(3) Functional deficits which require support
(4) Features that make the system user friendly to the cognitively impaired user, who may also have physical impairments.

# Appendix C    Other Scenarios Considered

In the process of developing a scenario for the Memex prototype, we brainstormed over several scenarios. In this section, we mention a few scenarios we considered before arriving at our final scenario.

## C.1   April, The Runner

April Greenside, a young woman in her late-twenties, has recently started running long distance races. She trains hard and tries to maintain a diet that will help her run longer and faster. April recently acquired a Personal Memex in the hope that it would help her in her new endeavor.

Today is another training day and April has to run an extra half-mile over her earlier running distance. As April gets ready for the day, she decides that she will update her Memex today to include some new information topics. Using the Memex's PC interface, she adds a new topic – "running trails". She then instructs the Memex to associate this topic with all race trails and training trails recorded by her SenseCam (which were annotated by her earlier), and related web pages and emails (by searching on some keywords). Then, she starts adding other information (attributes) to this new topic (possibly associating other information that has been recorded), like distance, difficulty, terrain features, location, splits. Having configured a new topic, April hopes to receive some good recommendations from her Memex that will help her in training.

April's checks her email to find a message from a fellow runner who has sent a link to an upcoming race. April quickly skims through race information on the page pointed to by the link. She marks the web page under the "races" topic of her Memex and knows that her Memex will capture all important information from the website (like trail-related and date information). On her request for a recommendation, the Memex suggests a list of practice trails that she may use to train for this race. Using historical information, the Memex also notifies her on expected weather on the day of the race. All this information will help April prepare for the race.

April now has just enough time to eat breakfast and then run for a doctor's appointment. She starts to drink milk and eat some fruit and cereal, when her Memex vibrates with a reminder message saying that the last time she ate fruit on the day of a training session, her stomach felt sick and she couldn't run beyond 1.5 miles. Then, using the Memex to recall the breakfast options suggested her coach, she finishes breakfast and heads out for the doctor's appointment.

April had complained about a mild continuous pain in the muscles around her knee and ankle the last time she finished running 6 miles. Today she is not feeling any pain. April's Memex (through the SenseCam) recorded information about this pain by recording voice annotations and taking pictures for last week, keeping track of the state of her knee and ankle. The doctor uses this information to advice April to perform certain stretching exercises before and after a training session and a race.

April now heads for the running trail, which is trail-park of several interconnecting trails. Since the trail park is not very well marked, April uses the GPS unit on her SenseCam and information from the Memex to guide her through her training session. April uses her SenseCam to record voice annotations and all other information as she is running through the practice trail. She knows she can use the Memex to recall this information for later use.

## C.2   Jay, The Musician

Jay is a drummer who is very serious about his music.  In addition to taking private lessons, he is a member of several groups, from a large orchestra to a jazz trio.  He spends hours each week practicing on his own, 2 hours each week in a private lesson and several hours in rehearsals.  He sets up his Memex to associate his different performances with the different groups and different types of music.  He finds it easiest to first recall the music by people he was with at the time and the composer and song title.  Jay wants to optimize his practice time.  First he uses his Memex to recall which sections of music his teacher focused on.  He replays what his teacher said, followed by the audio file of how the teacher demonstrated the music.  He practices those sections until he can play them correctly.  Later, during his practice time, he switches to his jazz performance.  He remembers that last week, during rehearsal, he did an impromptu drum solo that everyone loved and that they wanted to include in the next gig.  After the drum solo, he had reached in his pocket and pressed the '1' key on his cellphone, indicating to his Memex that the previous recording was very important.  Jay quickly retrieves the recording and reviews it several times until he knows it and can repeat it.

(The person with visual impairment has a hard time reading the music, so they review past performances or their teacher's demonstrations so that they can repeat the music by ear without needing the sheet music.)
(The user with MCI forgot what to practice, so he uses his Memex to review what the teacher or conductor said and which pieces of music were studied.)

## C.3   Track Meet

Dan and Ann go to their daughter's cross-country track meets every Saturday in the Fall.  After several meets, they know most of their daughter's teammates by name, but there are still several who are hard to remember.  In addition, much of the time at the meet is spent making small talk with the other parents.  Sometimes it's difficult to remember their names, because they are only encountered occasionally.  It would be especially helpful to remember the last conversation and maybe some topics of potential conversation.  For example, Dan remembers that one runner's mother works at the Cranwell Center for International Students and he would like to ask her how their fund-raising efforts for Tsunami victims are going.
The race then begins.   At this week's meet, there are several hundred runners in the race.  Dan and Ann can see their team, a blur of girls dressed in blue, run by.  It can be so difficult to tell the runners and teams apart.   They are wearing similar outfits and it seems like many teams, like Blacksburg, favor the royal blue color.

After a few minutes, the runners begin to spread out and it is now possible to see who is in the lead.  With hundreds of runners, it is impossible to remember the stats and times for other runners.  In addition, since every cross-country course is different, to truly rate the runners, Dan needs to know not only their most recent time for running 5 kilometers, but their time on similar courses in the past.

The lead runners start running towards the observation point where Dan is standing. His Memex shows him that the girl in first is from Cave Spring high school and her times from this course last year and her personal record for a 5K run. He sees that, even though she starts out fast, she usually finishes a little later in the race and feels that this runner, although she is currently in first place, is not a threat to his daughter's position.

After watching a few more runners and comparing their stats, Dan realizes that three of the top 15 runners are from the same team, a team with maroon uniforms that say 'Knights' on the front. What is this team and are they a threat to our team's position? Dan's Memex reports that the Knights are the team from Cave Spring High School, and even though they have three runners running well, their fourth and fifth runners historically do poorly. Since it takes five good runners to win the race as a team, Dan breathes a sigh of relief and thinks that he wants to pay more attention to who is in the fourth and fifth position for each team.

A girl from the home team is running by. Dan doesn't remember her name, but his Memex reminds him in time for Dan to yell, "Go Sarah". Her split times are looking good for the course; she ran the first two miles in less time than she has in the past. He cheers more enthusiastically.

Meanwhile, the coach is also watching the race. His Memex doesn't remind him of team names or individual's names; his own mind can recall that information immediately. More detailed split times and statistics show up on his head-up display.

After the race, he considers his team's performances. He asks them to submit the weekly analyses of their training. This includes a summary and analyses of their workout schedule (and if they really ran when they were supposed to), what they ate, how long before the race they ate, the amount of sleep they've had, especially for the days before the race and how much water they drank. One of his runners, Katie, had a stomach cramp after the race again today. He will ask her to run an analysis of her diet before her good races and before the races when she cramped up. His Memex notifies him that the official race results have been completed. He is pleased to confirm that his team has indeed won the race and he strides proudly over to the team to give them accolades for their great performance.

(Afterwards, he ponders what a great tool his Memex is, so he wants to tell his friend, the football coach, how he uses his Memex. It could be even more useful in football because it could help the players learn strategies for the game and the characteristics of the opposing team.)

# Appendix D    Interview Notes

## D.1    Bill Hobach and Hal Brackett, Assistive Technologies

We interviewed Bill Holbach and Hal Brackett on Monday, 12 October 2005.   Bill Holbach is the Assistive Technologies coordinator and Hal Brackett is the Special Services Manager.   The goal of the Assistive Technologies and Special Services groups are to provide equal access to all students at the University by providing various computer-related assistive technologies, such as screen magnifiers, voice recognition software and accessible workstations.   More details about Assistive Technologies can be found at http://www.it.vt.edu/organization/lt/assistive_technologies.html.

In their role providing services for students with a wide range of disabilities, Bill and Hal have both expertise and anecdotal evidence on what is needed and useful to the student population on campus.

We gave them a copy of the executive summary and then had a loosely structured conversation about the possibilities and constraints for a personal Memex.

One of the general concerns shared by caregivers is that of the governmental structures imposed on people with disabilities.  The Individuals with Disabilities Education Act (IDEA) provides for aides for students.  Once the students reach adulthood, the Americans with Disabilities Act (ADA) promotes a level of independence to which the students are not accustomed.  With the constant support of aides and caregivers, students can acquire 'learned helplessness,' which results in even less independence than the students are capable of.

Bill and Hal share a vision of encouraging independence for individuals with disabilities.  Hal envisions a device or system that he calls 'MOM', short for 'Micro Omnicient Manager'.  Your MOM would watch that you have what you need for the day and that you've gotten on the correct bus.  As you progress towards independence, it stops nagging you.

Our selected population groups with disabilities were individuals with mild cognitive impairment and individuals with macular degeneration.  Bill and Hal suggested that we focus on general needs, instead of a specific disease.  Instead of focusing on Macular Degeneration, which has several levels of vision degradation, we should focus on general visual impairment.   We then continued to discuss the specific conditions.

Macular Degeneration can range from very little visual impairment to complete blindness.  In general, users with Macular Degeneration see no color or detail.  The condition is similar to having night vision during the day and many individuals like to keep light levels low.  As an aside, Hal wondered how well the SenseCam would function in perpetually low-light conditions. Bill and Hal also reminded us to remember that the condition changes over time, so no single interface solution would be appropriate, even for a single user.

Mild Cognitive Impairment was discussed along with Learning Disabilities and other cognitive impairments.  Individuals with learning disabilities are easily distractible and need to focus.

Individuals with Obsessive Compulsive Disorder (OCD) need help getting out of the OCD loop and back on the task at hand. Students with Learning Disabilities need to learn the changes needed in their daily routine to succeed on campus. Studies have shown that failure in college for students with learning disabilities is not due to a lack of intelligence, but in problems with organization.

An individual with a Cognitive Disability (an IQ less than 100) can be successfully trained to automaticity, even with a complex task. Hal gave an anecdote about a gunner in the Vietnam War who had a Cognitive Disability. He could perfectly disassemble and reassemble a gun. However, in flight, he might not remember not to shoot the gun at a neighboring friendly aircraft.

Suggestions for the possible uses of a Memex for handicapped persons include:
- Guiding through an unfamiliar physical environment
- Queuing/Prioritizing information/tasks
- Helping them be independent. Once a task has been learned to automaticity, the Memex may stop explicitly reminding the user to perform the task. The Memex may continue to monitor the task to be sure that the user does not forget to perform the task under changing circumstances. For example, a notification for lunch could be modified to not remind the user to eat lunch, but to remind the user if it appears that he has forgotten to eat lunch. In addition to reminders to eat, other cues include reminders to catch the bus or to take medications.
- Monitor progress in tasks and make suggestions for achieving success. This monitoring function could also be done by a remote caregiver, who could then in turn, make suggestions to the student.
- Provide feedback to remote caregivers. This step provides support for the stage between having a caregiver constantly present to complete independence. A remote caregiver could monitor the user's progress in taking medicines or following a daily routine. The remote caregiver could also watch for behavior patterns which could signal a problem or situations that need interaction by the caregiver.

### D.1.1   Interface Features

Considerations

- Ability to do things quickly
- Physical dexterity and visual acuity so that they may see/handle the device properly
- Consistency in display items (should not change much for related items)
- Offer multiple modalities and let the users select which one they wish to use. For example, auditory cues may be useful for individuals with visual impairments, but frustrating to individuals who also have problems with short term memory.
- Bill suggested that instead of an all-purpose, general voice recognition system, the interface use a voice interface with a limited number of commands ("discrete voice recognition"). The user could say "help" to get a list of the possible commands.

Ideas
- OCD

- Auditory/Visual cues
- Sliding bars
- Choice of text/graphics

### D.1.2 Other Features

Considerations
- Use threshold settings to give reminders/recommendations.

Ideas
- Facial identification of select people they meet. Instead of implementing a global facial recognition system (which hasn't been shown to work), create a system to recognize the faces of people in the users' select groups. This would include the people they see routinely at school, at work or at the doctor's office.
- Use the SenseCam/Memex (maybe in the form of a joystick) as a visual guide rather than a cognitive guide (example, following a car to some place or getting there based on directions to that place).

### D.1.3 Devices that may be used/extended to act as the interface for the Memex

Considerations
- Ubiquitous device
- Choose adaptable devices that are easy to use
- Does the device require charging, how easy is it to perform this task? Maybe Wireless mechanisms may be used (waving the device or placing it in a dock to charge it)
- Some devices, such as the alarm watch made by Timex, include very small connectors for connecting the device to a PC. Someone with limited dexterity would be frustrated attempting to connect a small cable. A wireless interface would be more appropriate.

Ideas
- Specialized watch by Timex (text display, sound)
- PDA
- Tablet PC
- RFID enabled devices

### D.1.4 Miscellaneous

- Remember people/objects/events in context but not out of context. Can the Memex help?
- Looking at a learning disability may help.
- Find out what are the changes/challenges in a day that they need to struggle with.
- Is the Memex going to provide real-time feedback (warn before tripping over a stone) or is it going to be something based on history ("You tripped the last time you were at this place in this position, so be careful.")

The final question on our executive summary asked about novel uses of the Personal Memex. Several novel ideas were proposed:
- Check the weather predictions on the Internet and remind the user to bring an umbrella or dress appropriately.

- See lightning or include a lightning detector and warn user to get inside as soon as possible.
- Instead of navigation by a set of instructions ("Turn right, walk one block, turn left."), let the user follow the visual path left by the Personal Memex. For example the user would see the view that the Memex saw when last walking the path and would just follow it.
- How can an artist with Macular Degeneration, who can not see color or detail, create art? Is there software that announces color of an image?

## D.2  Karen Roberto

She was addressing the issue mostly from the perspective of an older population.

People with cognitive disabilities have difficulty recalling calendar and time events. Relevance of time factors seems to be affected by the age group under consideration also. For people in the old-old group, time seems to be less important. Their activities are not time driven. Relating events to time can be mentally taxing for them. They are more comfortable recalling visual images. So, visual images can be a good way to link events in the schema builder.

Since MCI is more related to short-term memory rather than long-term memory, older people with MCI (acquired) who are computer literate will still remember how to use a computer and other technologies that they were acquainted with. So I think that it'll be a good idea to concentrate on acquired MCI.

These people need constant reminders. They have difficulty remembering appointments, forgetting medications.

When designing for this population, it will be easier on them, if the numbers of steps are minimal. So it's better to have a small number of sweeping steps than a lot of smaller steps.

Some kind of prompts (cueing) can be of great help for them to accomplish tasks, while using something like a MEMEX.

MCI doesn't always change into dementia. This happens in only about 10-12 % cases.

This population seems to be bad with finding directions also.

It will be a good idea to audio reminders about things like taking medication. We should include the ability to go back and check (Did I take the medication?)

She was of the opinion that we should define exactly which class of MCI population we will be referencing, cause there will be differences.

There is still a controversy about diagnosis – its not in the DSM for example.
Briann, her research assistant will send us an EndNote lib of their references.

The most cited is Petersen, RC 1999 in Archives of Neurology vol 56 pp 303-?

which seems to focus on STM and WM deficiencies.
there  is also Royall who believes it is about executive function and frontal lobe problems

MCI patients have problems with recall and recognition, but can still function independently for the most part. About 10-20% continue to Alzheimers or Demetia.

Most of this information is anecdotal from the clinical research her group has been doing:

MCI patients have a number of problems with time including perception and instruments for management such as calendars. Numbers too are problematic. They do have a tendency to forget certain categories of things including: objects, appointments, recent actions (done something), and names.

For recall, it may be better to represent time graphically (like a timeline structure) versus numerically (like a table). Numbers are (anecdotally) problematic.

Barriers to use are the problem of patients acknowledging the condition (typically they have devised compensatory strategies) and form factor for accessing the system.

### D.3   Virginia Reilly, ADA Coordinator

Dr. Reilly was of the opinion that macular degeneration has more to do with sensory loss than loss of cognitive processing. MEMEX is primarily a memory aid.  So does this really fit in well?

Since we are dealing with a population that is visually impaired, we should be really concentrating on the interface design (view).

For people with macular degeneration, the following aspects have to be incorporated in the design.
-   lighting conditions: they prefer dim lit conditions than brightly lit conditions
-   Contrast is important for them. They prefer high contrast.

There are lots of assistive tools that can make life easier for people with visual impairment.
-   scan and read software
-   CCTV : not portable
-   Palm : portable
-   JAWS

There will be lot of differences in designing for this group depending on whether the visual impairment was acquired or if it was present at birth. If it's acquired then the person is going to have a well developed visual schema where as you won't have a developed visual schema in the other case.

The design should provide for visual input and output. There is text to audio conversion software.

There is a limit to the auditory memory. So the design should make sure that there's no overloading of any particular sense.

There is no one template that can be universally used. We should provide for lot of customization. The menu design should be such that to access something there shouldn't be lot of steps involved. The items should be fairly easy to access.

Compatibility with other assistive technologies that are currently available can be a big plus.